\documentclass[12pt,a4paper,onecolumn,draftcls]{IEEEtran}
\usepackage{epsfig,graphics,subfigure,psfrag,amsmath,cases}
\usepackage{latexsym,amssymb,amsmath,epsfig,subfigure,algorithm}
\usepackage{algorithmic,graphicx}
\usepackage{array}
\usepackage{color}
\usepackage{url}
\usepackage{scrtime}

\author{\authorblockN{ Derrick Wing Kwan Ng\authorrefmark{1}, Ernest S. Lo\authorrefmark{2}, and Robert Schober\authorrefmark{1}}\\
\authorrefmark{1}Institute for Digital Communications, Universit\"at Erlangen-N\"urnberg, Germany\\
\authorrefmark{2}Centre Tecnol\`{o}gic de Telecomunicacions de Catalunya - Hong Kong (CTTC-HK)\\
Email: kwan@lnt.de, ernest.lo@cttc.hk, schober@lnt.de \vspace*{-20mm}
}

\title{\vspace*{-6mm}Energy-Efficient Resource Allocation in  OFDMA Systems with Hybrid Energy Harvesting Base Station\thanks{This paper has been presented in part at the IEEE Global Communications Conference (Globecom 2012), Anaheim, California, USA. }}

{\newtheorem{Thm}{Theorem}}
\newtheorem{Lem}{Lemma}

\newtheorem{Remark}{Remark}

\begin{document}
\maketitle

\begin{abstract}
 We study resource allocation algorithm design for energy-efficient
communication in an orthogonal frequency division multiple
access (OFDMA) downlink network with hybrid energy harvesting base station (BS). Specifically, an energy harvester and a constant energy source driven by a non-renewable resource are used for supplying the energy required for system operation.  We first consider
 a deterministic \emph{offline} system setting. In particular, assuming  availability of non-causal knowledge about energy arrivals and channel gains,  an \emph{offline} resource allocation problem is formulated  as a non-convex optimization problem over a finite horizon taking into account the circuit energy consumption, a finite energy storage capacity,  and a minimum required data rate.   We transform this non-convex optimization problem into a
convex optimization problem by applying time-sharing and exploiting the properties of non-linear
fractional programming  which results in an efficient asymptotically \emph{optimal offline} iterative
resource allocation algorithm for a sufficiently large number of subcarriers. In each iteration, the transformed
problem is solved by using Lagrange dual decomposition. The obtained resource allocation policy maximizes the weighted energy efficiency of data
transmission (weighted bit/Joule delivered to the receiver). Subsequently, we focus on  \emph{online} algorithm design.
A conventional stochastic dynamic programming approach is employed to obtain the \emph{optimal online} resource allocation algorithm which entails a prohibitively high complexity. To strike a balance between system performance and computational complexity,   we propose a low complexity \emph{suboptimal
online iterative algorithm} which is motivated by the \emph{offline} algorithm. Simulation results
illustrate that the proposed suboptimal online iterative resource allocation algorithm
does not only converge in a small number of iterations, but also
achieves a close-to-optimal system energy efficiency by utilizing only causal channel state and energy arrival information.
\end{abstract}

\vspace*{-2mm}
\begin{keywords}\vspace*{-2mm} Energy harvesting, green communication, non-convex optimization, resource allocation.
\end{keywords}

\vspace*{-3mm}
\section{Introduction}
\label{sect1}
Orthogonal frequency division multiple access (OFDMA) is  a viable
multiple access scheme for spectrally efficient communication
systems due to its flexility in resource allocation  and ability to
exploit multiuser diversity \cite{JR:OFDMA,CN:large_subcarriers}. Specifically, OFDMA converts a wideband
channel into a number of orthogonal narrowband subcarrier channels
and multiplexes the data of multiple users on different subcarriers.
In a downlink OFDMA system, the maximum system throughput can be
achieved by selecting the best user on each subcarrier and adapting
the transmit power over all subcarriers using water-filling. On the other hand, the increasing interest in high data rate services such as
video conferencing and online high definition video streaming has
led to a high demand for energy. This  trend has  significant financial
implications for service providers due to the rapidly increasing
cost of energy. Recently,  driven by environmental concerns,  green communication has received considerable interest from both industry
and academia \cite{Magazine:green1}-\nocite{Magazine:green2,Magazine:green3}\cite{CN:green_statistic}. In fact, the cellular networks consume world-wide approximately 60 billion
kWh per year. In particular, 80\% of the electricity in cellular networks is consumed by the base stations (BSs) which produce over a hundred million tons of
carbon dioxide per year  \cite{CN:green_statistic}. These figures are projected to double by the year 2020 if
no further actions are taken.  As a
 result,
 a tremendous number of green technologies/methods have been proposed   in the literature
 for maximizing the energy efficiency  (bit-per-Joule) of wireless communication systems  \cite{CN:ee}-\nocite{JR:energy_efficient_2,JR:energy_efficient_3}\cite{JR:limited_backhaul}.
In \cite{CN:ee}, a closed-form power allocation solution was derived for maximizing the energy efficiency of a point-to-point single carrier system with a minimum average throughput requirement. In \cite{JR:energy_efficient_2}-\cite{JR:limited_backhaul}, energy efficiency has been studied in cellular multi-carrier multi-user systems for  both uplink and downlink communications. Specifically,  in \cite{JR:energy_efficient_2}-\cite{JR:limited_backhaul}, the existence of a unique global maximum for the energy efficiency  was proven for different systems  and can be achieved by corresponding resource allocation algorithms. On the other hand, there have been recent research efforts to enhance the system energy efficiency by using multiple antennas \cite{CN:energy_efficient_2}-\nocite{JR:ee_letters}\cite{JR:TWC_large_antennas}.
In \cite{CN:energy_efficient_2} and \cite{JR:ee_letters}, power loading
algorithms were designed to maximize the energy efficiency of systems with collocated and distributed antennas, respectively. In \cite{JR:TWC_large_antennas}, the authors studied the energy efficiency of cellular networks with a large number of transmit antennas in OFDMA systems. Yet,  \cite{CN:ee}-\cite{JR:TWC_large_antennas} require the availability of an  ideal power supply such that a large amount of energy can be continuously used for system operations whenever needed. In practice,  BSs may not be connected to the power grid, especially in developing countries. Thus, the assumption of a continuous energy supply  made in \cite{CN:ee}-\cite{JR:TWC_large_antennas} is overly  optimistic in this case. Although these BSs can be possibly powered by diesel generators \cite{Mag:off-grid}, the inefficiency of diesel fuel power generators and high transportation costs of diesel fuel are obstacles for the provision of wireless services  in remote areas \cite{CN:off-grid}.  In such situations,
 energy harvesting is
particularly appealing since BSs can harvest energy from natural renewable energy sources such as solar, wind, and geothermal heat,  thereby reducing substantially the operating costs of the service providers. As a result, wireless networks  with energy harvesting BSs are not only envisioned to  be energy-efficient in providing ubiquitous service coverage, but also to be self-sustained.

The introduction of energy harvesting capabilities
 for BSs poses many interesting  new challenges for resource allocation algorithm design due to the time varying availability of the energy generated from renewable energy sources. In  \cite{JR:harvesting_single_user} and \cite{JR:harvesting_single_user3}, optimal packet scheduling and power allocation algorithms
were proposed for energy harvesting systems for minimization of the transmission completion time, respectively. In \cite{JR:harvesting_single_user2} and  \cite{JR:JSAC_Optimal_Policies}, the authors proposed optimal power control time sequences for maximizing the throughput by a deadline with a single energy harvester.
However, these works assumed a point-to-point narrowband communication system and the obtained results may not be applicable to
the case of wideband multi-user systems.
   In \cite{JR:harvesting_multi_user2}-\nocite{JR:harvesting_multi_user3}\cite{JR:harvesting_multi_user},  different optimal packet scheduling algorithms were proposed for additive white Gaussian noise (AWGN) broadcast channels for a set of preselected users. However, wireless communication channels are not only impaired by AWGN but also degraded by multi-path fading. In addition, dynamic user selection is usually performed to enhance the system performance.   On the other hand, although the amount of renewable energy is potentially unlimited,
the intermittent nature of energy generated by a natural energy source results in a highly random energy availability at the BS.
 For example,  solar energy
and wind energy are  varying significantly over time
due to weather and climate conditions. In other words, a BS powered solely by an energy harvester
 may not be able to maintain a stable operation and to guarantee a certain quality of service (QoS). Therefore,
a hybrid energy harvesting system design, which uses different energy sources in a  complementary  manner, is preferable in practice for providing uninterrupted service
\cite{JR:huawei,JR:hybrid_energy}. However,  the results in the literature, e.g.  \cite{CN:ee}-\cite{JR:harvesting_multi_user},  are only valid for systems with a single energy source and are not applicable to communication networks employing hybrid energy harvesting BSs.

In this paper, we address the above issues and focus on  resource allocation algorithm design for hybrid energy harvesting BSs. In Section \ref{sect:OFDMA_AF_network_model}, we introduce the adopted OFDMA channel model and hybrid energy source model. In Section \ref{sect:3}, we
formulate  offline resource allocation as an optimization problem by assuming non-causal knowledge of the channel gains and energy arrivals  at the BS. The optimization problem is solved via fractional programming and Lagrange dual decomposition which leads to an efficient iterative resource allocation algorithm. The derived offline solution serves as a building block for the design of a practical close-to-optimal online resource allocation algorithm in Section \ref{sect:4}  which  requires only causal knowledge of the channel gains and energy arrivals.  In Section \ref{sect:5}, we show that the proposed suboptimal algorithm does not only have a fast convergence, but also achieves a close-to-optimal performance.
\begin{figure}[t]\vspace*{-1.3cm}
\centering
\includegraphics[width=4.2in]{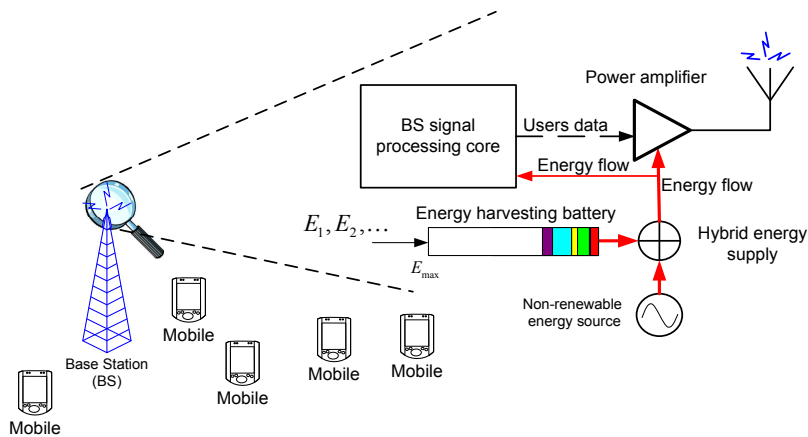}\vspace*{-3mm}
\caption{An OFDMA system  with a hybrid energy harvesting base station (BS) for signal transmission. Two energy sources are implemented in the system, i.e.,  a renewable energy harvesting source and a non-renewable energy source. \vspace*{-10mm}  }\label{fig:system_model}
\end{figure}\vspace*{-0.3cm}
\section{OFDMA System Model}
\label{sect:OFDMA_AF_network_model}

\vspace*{-4mm}
\subsection{Notation}
 A complex Gaussian random variable with mean $\mu$ and variance
$\sigma^2$ is denoted by ${\cal CN}(\mu,\sigma^2)$, and $\sim$ means
``distributed as".
$\big[x\big]^+=\max\{0,x\}$. $\big[x\big]^a_b=a,\ \mbox{if}\ x>a,\big[x\big]^a_b=x,\mbox{ if}\ b\le x\le a,\big[x\big]^a_b  =b,\ \mbox{if}\ b>x$. ${\cal E}_{\mathbf{x}}\{\cdot\}$ denotes statistical
expectation with respect to (w.r.t.) random variable $\mathbf{x}$.
\vspace*{-4mm}

\subsection{OFDMA Channel Model} \label{sect:channel_model}
 We consider an OFDMA
network which consists of a BS
and $K$ mobile users. All transceivers are equipped with a single antenna, cf. Figure
\ref{fig:system_model}. The total
 bandwidth of the system is $\cal B$ Hertz and there are $n_F$ subcarriers. The transmission time is $T$ seconds. We assume that the BS adapts the resource allocation policy (i.e., the power allocation and subcarrier allocation policies) $L$ times for a given period $T$.  The optimal value of $L$ and the time instant of each adaption will be provided in the next section. The downlink symbol received at user $k\in\{1,\,\ldots,\,K\}$ from the BS on subcarrier
$i\in\{1,\,\ldots,\,n_F\}$ at time instant\footnote{In practical systems, the length of the cyclic prefix of an orthogonal frequency-division multiplexing (OFDM) symbol is chosen to be larger than the root mean square delay spread of the channel. Nevertheless, the channel gains from one subcarrier to the next may change considerably. Therefore, we use a discrete model for the frequency domain. On the other hand, the coherence time for a low mobility user is about $200$ $ms$ and an OFDM symbol in Long-Term-Evolution (LTE) systems has a length of 71.3 $\mu s$. Thus,  during a transmission time $T$ much longer than the coherence time, e.g., $T\gg$ 200 $ms$, a few thousands of OFDM symbols are transmitted. Therefore, we use a continuous time domain signal model for representing the time variation of the signals.    } $t$,  $0\le t\le T$, is given by
\begin{eqnarray}
y_{i,k}(t)&=&\sqrt{P_{i,k}(t)g_{k}(t)}{H}_{i,k}(t) x_{i,k}(t)+z_{i,k}(t),
\end{eqnarray}
where $x_{i,k}(t)$ is the symbol transmitted from the BS to user $k$ on
subcarrier $i$ at time $t$.
$P_{i,k}(t)$ is the transmit power for the link between the BS and user $k$ on subcarrier $i$. ${H}_{i,k}(t)$ is the small scale fading coefficient between the BS and user
$k$ on subcarrier $i$ at time $t$.
$g_k(t)$ represents the joint effect of path loss and shadowing
between the BS and user $k$ at time $t$. $z_{i,k}(t)$ is the AWGN in subcarrier $i$ at user $k$ with
distribution ${\cal CN}(0,N_0)$, where $N_0$ is the noise power
spectral density.\vspace*{-4mm}
\subsection{Models for Time Varying Fading and Energy Sources} \label{sect:channel_model}
\begin{figure}[t]\vspace*{-1.25cm}
\hspace*{-1cm}\centering
\includegraphics[width=5.35in]{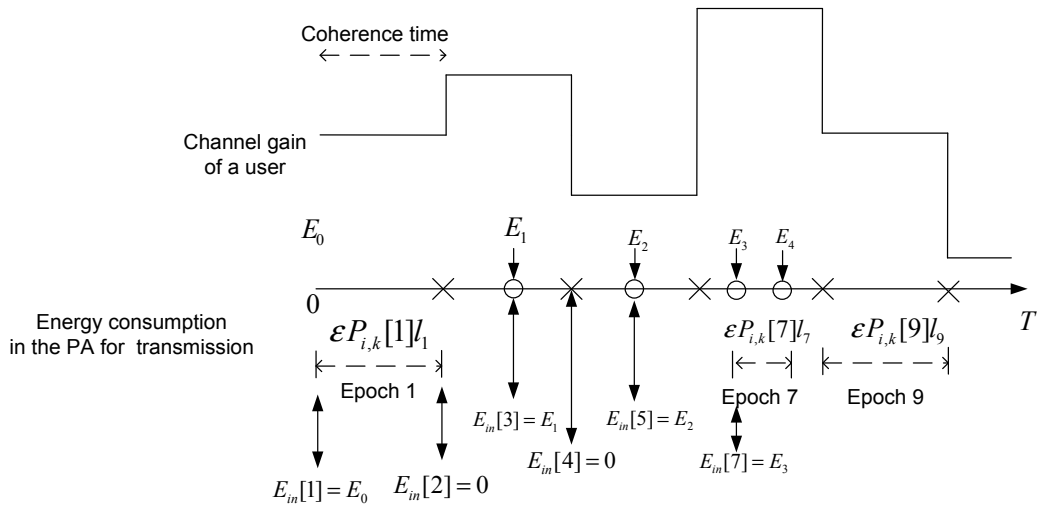}\vspace*{-4mm}
\caption{An illustration of epoch and $E_{in}[\cdot]$ in (\ref{eqn:cap_discrete}) for different events at different  arrival times.   Fading changes and  energies are harvested at time instants denoted by  $\times$ and $\circ$,  respectively.  }\vspace*{-1mm}\label{fig:epoch}\vspace*{-8mm}
\end{figure}

In the BS, there are two energy sources for supplying the energy  required for system operation, i.e., an energy harvester and a constant energy source driven by a non-renewable resource, cf. Figure \ref{fig:system_model}.   In practice, the  model for the energy harvester  depends on its specific
implementation. For instance, both solar panel  and
wind turbine-generator  are able to generate renewable energy for communication purposes. Yet, the energy harvesting characteristics (i.e., energy arrival dynamics and the amount of energy being harvested) are different  in both cases. In order to
provide a general model for energy harvesting communication systems, we do not assume a
particular type of energy harvester. Instead, we
 model the energy output characteristic of the energy harvester  as a stochastic process in order to isolate the considered problem from specific
implementation assumptions. In particular, we adopt a similar system model as in \cite{JR:JSAC_Optimal_Policies} for
modeling the time varying nature of the communication channels and the random energy arrival behavior in the energy harvester.
We assume that the energy arrival times in
the energy harvester are modeled as a Poisson counting process with rate $\lambda_E$.
Therefore, the energy arrivals
 occur in a countable number of time instants, which
are indexed as  $\{t_1^E,t_2^E,\ldots\}$. The inter-occurrence time
between two consecutive energy arrivals, i.e., $t_{b}^E-t_{b-1}^E,
b\in\{1,2,\ldots\}$, is exponentially distributed with mean
$1/\lambda_E$. Besides, $E_b$ units of energy
arrive (to be harvested) at the BS at  time $t_b^E$. On the other hand, a block fading
communication channel model is considered, cf.
\cite{JR:JSAC_Optimal_Policies,JR:block_fading,CN:block_fading}.
Without loss of generality, we denote the  time instants where the fading level changes as $\{t_1^F,t_2^F,\ldots\}$. We note that the
fading level in $0<t\le t_1^F $ is constant but changes\footnote{In the paper, the changes in channel gain
refer to the changes in $H_{i,k}(t)$.  In fact,
$g_{k}(t)$ is assumed to be a constant over time $T$. The coherence time of the shadowing
and path loss is proportional to the coherence distance. For
instance, the coherence distance in a suburban area is around
100-200 m  and tens of meters in an urban area \cite{JR:shadow}.
Assuming a coherence distance of 100 m, this results in a coherence
time of about 120 seconds at 3 km/h (pedestrian speed). Then, the
coherence time of the multipath fading is an order of magnitude
smaller than the coherence time of the shadowing.} to an
independent value in the next time interval of fading block, $t_1^F<t\le t_2^F$, and
so on.  The length of each fading block is approximately equal to the
coherence time of the channel \footnote{In the paper, we assume that
all users have a similar velocity such that their channels have a
similar coherence time. Yet, our model can be generalized to
different coherence times for different users, at the expense of a more
involved notation.}, cf. Figure \ref{fig:epoch}. The incoming energy is collected by an energy harvester
and is buffered in the battery
before it is used for data transmission.  On the other hand, we
assume that $E_0$ units of energy arrive(/are available) in the battery at $t_0^E=0$ and the maximum amount of  energy storage in the battery is denoted by $E_{\max}$. In the following, we refer to a change of the channel gain of any user or the energy level in the battery as an
\emph{event} and the time interval between two consecutive events as an \emph{epoch}. Specifically, \emph{epoch}
$l$, $l\in\{1,2,\ldots\}$, is defined as the time interval $[t_{l-1}, t_{l})$, where $t_{l-1}$ and $t_{l}$ are the time instants at which successive
events happen, cf. Figure \ref{fig:epoch}.

\begin{Remark}
 We note that the major assumption made in the modelling of the problem  is the  stationarity  and ergodicity of the fading and the energy arrival random processes. In fact, the assumption of particular distributions for the changes in fading gains, time of changes in fading gains, and/or energy arrival times   do not change the structure of the algorithms presented in the paper as long as the corresponding distributions are known at the BS. This knowledge can be obtained via long term measurements. The assumption of a Poisson counting process for the energy arrivals is made for illustration of the countability of the incoming energy arrivals.

\end{Remark}

In the considered model, the transmitter
can draw the energy required for  signal transmission and signal processing from both the battery\footnote{Note that the term ``battery"
is used interchangeably with the term ``energy harvester" in the paper. }  and the traditional energy source.
In particular,  the instantaneous total radio frequency (RF) transmit power of the power amplifier (PA) for user $k$ in subcarrier $i$  at time instant $t$ can be modeled as
\begin{eqnarray}
 P_{i,k}(t) =P_{i,k}^E(t)+P_{i,k}^N(t),\quad\forall i,k,0\le t\le T,
\end{eqnarray}
 where  $P_{i,k}^E(t)$ and $P_{i,k}^N(t)$ are the  portions of the instantaneous transmitted power taken from the energy harvester and the non-renewable energy source for user $k$ in subcarrier $i$ at time instant $t$, respectively.  Furthermore,  we model the energy consumption required for signal processing as
  \begin{eqnarray}
 \int_0^t \Big(P_C^E(u)+P_C^N(u)\Big)\, du=P_C t,\quad 0\le t\le T,
 \end{eqnarray}
 where  $P_C^E(t)$ and $P_C^N(t)$ are the portions of the instantaneous power required for signal processing drawn from the energy harvester and the non-renewable energy source, respectively. $P_C$ is the required constant  signal processing power at each time instant and includes the power dissipation in the mixer, transmit filters,
frequency synthesizer, and  digital-to-analog converter (DAC), etc.

 Since two energy sources are implemented at the BS, we have to consider the physical constraints imposed by both  energy sources, which are described in the following.

\subsubsection{Energy Harvesting Source}

There are two inherent constraints on the energy harvester:
\begin{eqnarray}\label{eqn:energy_harvester_causality}
\mbox{C1:}\underbrace{\sum_{i=1}^{n_F}\sum_{k=1}^K\int_0^{t^E_b -\delta} \varepsilon  s_{i,k}(u)P_{i,k}^E(u)\, du}_{\mbox{Energy  from energy harvester used in PA }}\hspace*{-3mm}+\int_0^{t^E_b -\delta} P_C^E(u)\, du \hspace*{-1mm}&\le& \hspace*{-1mm}\sum_{j=0}^{b-1}E_j,\, \forall b\in\{1,2,\ldots\},\end{eqnarray}\begin{eqnarray}
\label{eqn:energy_harvester_max}
\mbox{C2:} \sum_{j=0}^{d(t)}E_j-\sum_{i=1}^{n_F}\sum_{k=1}^K\int_0^{t} \varepsilon  s_{i,k}(u)P_{i,k}^E(u)\, du-\int_0^t P_C^E(u)\, du&\le& E_{\max}, \quad  0\le t\le T,
\end{eqnarray}
where  $\delta\rightarrow 0$ is an infinitesimal positive constant for modeling purpose\footnote{The integral in equation (\ref{eqn:energy_harvester_causality}) is defined over a half-open interval in $[0, t^E_b)$. As a result, the variable $\delta$ is used to account for an infinitesimal gap between the upper limit of integration  and the boundary $t^E_b$. }, $d(t)=\arg \underset{a}{\max}\{t_a^E: t_a^E\le t\}$, $s_{i,k}(t)\in\{0,1\}$ is the binary subcarrier allocation indicator at time $t$,
 and  $\varepsilon\ge 1$ is
constant which accounts for the inefficiency of the PA.
For example, when $\varepsilon=10$, 100 Watts of power are consumed in the PA   for every 10 Watts of
power radiated in the RF. In other words, the power efficiency is
$\frac{1}{\varepsilon}=\frac{1}{10}=10\%$. Constraint C1 implies that in every time instant, if the BS draws energy from the energy harvester to cover the energies required at the PA and for signal processing, it is constrained to use at most the amount of stored energy currently available (causality), even though more energy may possibly arrive in the future. Constraint C2 states that the energy level in the battery never exceeds $E_{\max}$ in order to prevent energy overflow in(/overcharging to) the battery. In practice, energy overflow may occur if the BS is equipped with a  small capacity battery.
\subsubsection{Non-renewable Energy Source}
In each time instant, a maximum power of $P_N$ Watts can be provided by the non-renewable energy source
to the BS. In other words, a maximum of $P_N t$ Joules of energy can be drawn from the non-renewable source from time zero up to time $t$. As a result, we have the following constraint on drawing power/energy from the non-renewable energy source at any time instant:
\begin{eqnarray}
\mbox{C3:}&&\underbrace{ \sum_{i=1}^{n_F}\sum_{k=1}^K \varepsilon s_{i,k}(t)P_{i,k}^N(t)}_{\mbox{Power from non-renewable source used in PA }}+ P_C^N(t) \le P_N ,\quad  0\le t\le T.
\end{eqnarray}


\section{Offline Resource Allocation and Scheduling Design}\label{sect:3}
In this section, we design an offline resource allocation algorithm  by assuming the availability of  non-causal knowledge of energy arrivals and channel gains.\vspace*{-4mm}
\subsection{Channel Capacity and Energy Efficiency}
\label{subsect:Instaneous_Mutual_information}
In this subsection, we define the adopted system performance
measure. At the BS, the data buffers for the users are assumed to be always full and there are no empty
scheduling slots due to an insufficient number of data packets at
the buffers. Given perfect channel state information (CSI) at the receiver, the channel
capacity\footnote{In general, if the future CSI is not available at the BS, the randomness of the multipath
fading causes resource allocation  mismatches at the BS
which decreases the system capacity.  For instance, if  only causal
knowledge of multipath coefficients is available at the BS,  the BS may transmit exceedingly large amounts of
power at a given time instant and exhaust all the energy of the
energy harvester, even though there may be much better channel
conditions in the next fading block which deserve more transmission energy
 for improving the system energy efficiency/capacity.} between the BS and user $k$ on subcarrier $i$ over a transmission period of $T$ second(s) with
subcarrier bandwidth $W=\frac{\cal B}{n_F}$ is given by
\begin{eqnarray}\label{eqn:cap}
&&C_{i,k}=\int_{0}^T s_{i,k}(t)W\log_2\Big(1+P_{i,k}(t)\Gamma_{i,k} (t)\Big) dt\quad \mbox{    where    } \quad \Gamma_{i,k}(t)=\frac{g_k(t)|H_{i,k}(t)|^2}{N_0W}.
\end{eqnarray}

The \emph{weighted total system capacity} is defined as the weighted sum of the total number
of
 bits successfully delivered to the $K$ mobile users over a duration of $T$ seconds and is given by
\begin{eqnarray} \hspace*{-5mm}
 \label{eqn:avg-sys-goodput} && \hspace*{-5mm} U({\cal P},  {\cal S})=\sum_{k=1}^K
 \alpha_k\sum_{i=1}^{n_F}
C_{i,k},
\end{eqnarray}
where ${\cal
P}=\{ P_{i,k}^E(t), P_{i,k}^N(t),P_C^E(t), P_C^N(t), \forall
i,k, 0\le t\le T\}$ and ${\cal{S}}=\{s_{i,k}(t), \forall
i,k, 0\le t\le T\}$ are the power and subcarrier allocation policies,
respectively. $0<\alpha_k\le 1$ is a positive constant provided by  upper layers, which allows the BS to give different
priorities to different users and to enforce certain notions of
fairness. On the other hand, we take into account the total energy consumption of the system by including it in  the optimization objective
function.   For this purpose, we model the weighted \emph{energy dissipation} in the system as the
sum of two dynamic terms
\begin{eqnarray}
 \label{eqn:power_consumption}
 U_{TP}({\cal P}, {\cal S})= \int_0^T \Big(  \phi P_C^E(t)+P_C^N(t)\Big)\, dt +
 \sum_{k=1}^K \sum_{i=1}^{n_F}
  \int_{0}^T s_{i,k}(t)\varepsilon\Big( \phi P_{i,k}^E(t) + P_{i,k}^N(t)\Big)\, dt,
\end{eqnarray}
where $\phi$ is a positive constant imposed on the use of the harvested energy. The value of $\phi$ can reflect either a normalized physical
cost (e.g., relative cost for  maintenance/operation of both sources of energy) or a normalized virtual cost (e.g., energy usage preferences),  w.r.t. the usage of the non-renewable energy source \cite{JR:virtual_cost}. In practice,  we set $0< \phi < 1$ to encourage  the BS to consume energy from the energy harvesting source.
The first term  and second term in (\ref{eqn:power_consumption})
denote the total weighted  energy consumptions in the signal processing unit and the PA, respectively. Hence, the \emph{weighted energy efficiency} of the considered
system over a time period of $T$ seconds is defined as the total average number of weighted bit/Joule
\begin{eqnarray}
 \label{eqn:avg-sys-eff} \hspace*{-8mm}U_{eff}({\cal P}, {\cal
 S})&=&\frac{U_{}({\cal P}, {\cal S})}{U_{TP}({\cal P},{\cal
 S})}.
\end{eqnarray}
\vspace*{-8mm}
\subsection{Optimization Problem Formulation}
\label{sect:cross-Layer_Formulation}

The optimal power allocation policy, ${\cal P}^*$, and subcarrier allocation policy, ${\cal S}^*$, can be
obtained by solving
\begin{eqnarray}
\label{eqn:cross-layer}&&\hspace*{30mm} \max_{{\cal P},  {\cal S}
}\,\, U_{eff}({\cal P}, {\cal S}) \\
\notag \hspace*{20mm}\mbox{s.t.} &&\hspace*{28mm}\quad\mbox{      C1, C2, C3} \nonumber\\
&&\hspace*{-35mm}\notag\mbox{C4:} \int_0^t \Big (P_C^E(u)+P_C^N(u)\Big)\, du=P_C t,\, 0\le t\le T, \quad\quad\mbox{C5:}\sum_{k=1}^K
\sum_{i=1}^{n_F}
C_{i,k} \ge R_{\min},\\
&&\hspace*{-35mm}\notag\mbox{C6:}\sum_{i=1}^{n_F}\sum_{k=1}^K P_{i,k}(t)s_{i,k}(t)  \le P_{\max}, \, 0\le t\le T,
\hspace*{16mm}\mbox{C7:}\,\, s_{i,k}(t)=\{0,1\}, \,\, \forall i,k,0\le t\le T,\quad\quad \\
&&\hspace*{-35mm}  \mbox{C8:}\,\, \sum_{k=1}^K s_{i,k}(t)\le 1, \,\, \forall i,0\le t\le T, \quad\mbox{C9: } P_{i,k}^N(t),P_{i,k}^E(t),P_{C}^N(t),P_{C}^E(t)\ge 0,\, \quad\forall i,k,0\le t\le T,\notag
\end{eqnarray}
where C4 ensures that the  energy required for signal processing is always available\footnote{ In the considered
system, the BS always has sufficient
energy for  CSI estimation despite the intermittent nature of energy generated by the energy harvester. Indeed, the BS is able to extract energy from
both the traditional power supply (from the power generator)
and the energy harvester. In the worst case, if the energy harvester
is unable to harvest enough energy from the environment, the BS
can always extract power from the traditional power supply  for supporting   the
energy consumption of signal processing in the BS. }.  C5 specifies the
minimum system data rate requirement $R_{\min}$ which acts as a QoS constraint for the system. Note that although
variable $R_{\min}$ in C5 is not an optimization variable in this
paper, a balance between energy efficiency and aggregate system
capacity can be struck by varying $R_{\min}$.   C6 is a constraint on the maximum transmit power of the BS. The value of $P_{\max}$ in C6 puts a limit on the transmit spectrum mask to control the amount of out-of-cell
interference in the downlink at every time instant.
Constraints C7 and
C8 are imposed to guarantee that each subcarrier will be used to serve at most one
user at any time instant. C9 is the non-negative constraint on the power allocation variables.

\begin{Remark}
 We note that an individual data
rate requirement for each user can be incorporated into the current problem formulation by  imposing the individual data rate requirements as additional constraints in the problem formulation \cite{JR:Meixia,JR:individual_data_Rate}, i.e., $\mbox{C10}: \sum_{i=1}^{n_F}
C_{i,k} \ge R_{\min_k}, {k\in \cal D}$, where ${\cal D}$  is a set of \emph{delay sensitive} users and
$R_{\min_k}$ is a constant which specifies the minimal required data
rate of user $k$. The resulting problem can be solved via a similar approach as used for solving the current problem formulation.
\end{Remark}
\vspace*{-4mm}
\subsection{Transformation of the Objective Function} \label{sect:solution_dual_decomposition}

The optimization problem in (\ref{eqn:cross-layer}) is non-convex due to the
fractional form of the objective function and the combinatorial constraint C7 on the subcarrier allocation variable. We note that there
is no standard approach for solving non-convex optimization
problems. In order to derive an efficient power allocation algorithm
for the considered problem, we introduce a
transformation to handle the objective function via nonlinear fractional programming \cite{JR:fractional}.
  Without loss of generality, we define the
maximum weighted energy efficiency $q^*$ of the considered system as
\begin{eqnarray}
q^*=\frac{U({\cal P^*},{\cal S^*})}{U_{TP}({\cal
P^*}, {\cal
 S^*})}=\max_{{\cal P}, {\cal S}}\,\frac{U({\cal P}, {\cal
S})}{U_{TP}({\cal P}, {\cal
 S})}.
\end{eqnarray}
We are now ready to introduce the following Theorem.
\begin{Thm}\label{Thm:1}
The maximum weighted energy efficiency $q^*$  is achieved \emph{if and only
if}
\begin{eqnarray}\label{eqn:penalty_q}
\max_{{\cal P},  {\cal S}}&& \hspace*{-2mm}\,U({\cal
P}, {\cal S})-q^*U_{TP}({\cal P}, {\cal
 S})=U({\cal
P^*}, {\cal S^*})-q^*U_{TP}({\cal P^*}, {\cal
 S^*})=0,
\end{eqnarray}
\end{Thm}
for $U({\cal P}, {\cal S})\ge0$ and $U_{TP}({\cal
P}, {\cal S})>0$.

 \emph{\quad Proof:} The proof of Theorem 1 is similar to the proof in \cite[Appendix A]{JR:TVT_secrecy_kwan}.

\hspace*{-6mm} Theorem \ref{Thm:1} reveals that for any objective function in fractional form, there exists an
equivalent
objective function in subtractive form,  e.g. $U({\cal P},
{\cal S})-q^*U_{TP}({\cal P}, {\cal S})$ in the
considered case, which shares the same optimal resource allocation policy. As a result, we can focus on the equivalent
objective function for finding the optimal offline resource allocation policy  in the rest of the paper.

\begin{table}\caption{Iterative Resource Allocation Algorithm for Suboptimal Online and Optimal Offline Designs.}\label{table:algorithm}\vspace*{-1cm}
\small
\begin{algorithm} [H]                    
\caption{Iterative Algorithm for Suboptimal Online and Optimal Offline Designs}          
\label{alg1}                           
\begin{algorithmic} [1]
\small          
\STATE Initialize the maximum number of iterations $I_{max}$ and the
maximum tolerance $\Delta$
 \STATE Set maximum energy
efficiency $q=0$ and iteration index $n=0$

\REPEAT [Main Loop]
\IF {\{Offline Problem\}}

 \STATE  Solve the inner loop problem in ($\ref{eqn:inner_loop}$)  for
a  given $q$ and obtain resource allocation policies $\{{\cal P'}, {\cal S'}\}$
 \ELSE   [Online Problem]
  \STATE Solve the inner loop problem in ($\ref{eqn:lag_offline}$)  for
a  given $q$ for each event and obtain resource allocation policies $\{{\cal P'}, {\cal S'}\}$

 \ENDIF

\IF {$U({\cal P'}, {\cal S'})-q U_{TP}({\cal P'}, {\cal
 S'})<\Delta$} \STATE  $\mbox{Convergence}=\,$\TRUE \RETURN
$\{{\cal P^*,S^*}\}=\{{\cal P',S'}\}$ and $q^*=\frac{U({\cal
P'}, {\cal S'})}{ U_{TP}({\cal P'}, {\cal
 S'})}$
 \ELSE \STATE
Set $q=\frac{U({\cal P'}, {\cal S'})}{ U_{TP}({\cal
P'}, {\cal
 S'})}$ and $n=n+1$ \STATE  Convergence $=$ \FALSE
 \ENDIF
 \UNTIL{Convergence $=$ \TRUE $\,$or $n=I_{max}$}

\end{algorithmic}
\end{algorithm}\vspace*{-1.6cm}
\end{table}

\vspace*{-4mm}
\subsection{Iterative Algorithm for Energy Efficiency Maximization} \label{sect:algorithm}
In this section, an iterative algorithm (known as the
Dinkelbach method \cite{JR:fractional})  is proposed for solving
(\ref{eqn:cross-layer}) by exploiting objective function $ \,{U}({\cal P}, {\cal S})-q{U}_{TP}({\cal
P}, {\cal
 S})$. The
proposed algorithm is summarized in Table \ref{table:algorithm}. Its convergence to the optimal energy efficiency is guaranteed if we are able to solve the inner problem (\ref{eqn:inner_loop}) in each iteration.

\emph{\quad Proof: }Please refer to \cite[Appendix B]{JR:TVT_secrecy_kwan}  for a proof of convergence.

 As shown in Table
\ref{table:algorithm}, in each iteration of the main loop, we solve
the following optimization problem for a given parameter $q$:
\begin{eqnarray}\label{eqn:inner_loop}
&&\hspace*{-1mm}\max_{{\cal P}, {\cal S}}
\quad\,{U}({\cal P}, {\cal S})-q{U}_{TP}({\cal
P}, {\cal
 S})\nonumber\\
&&\hspace*{-10mm}\mbox{s.t.} \,\,\mbox{C1, C2, C3, C4, C5, C6, C7, C8, C9}.
\end{eqnarray}

\subsubsection*{Solution of the Main Loop Problem}
Although the objective function is now transformed into a subtractive form which is easier to handle, there are still two obstacles in  solving the above problem.
First, the equivalent problem in each iteration is a mixed combinatorial and convex
optimization problem. The combinatorial nature comes from the binary
constraint C7 for subcarrier allocation. To obtain an optimal solution,
an exhaustive search is needed in every time instant which entails a complexity  of ${\cal O}(K^{n_F})$ and is computationally infeasible for $ K,n_F \gg 1$.
Second, the optimal resource allocation policy is expected to be time varying in the considered duration of $T$ seconds. However, it is unclear how often the BS should update the resource allocation policy which is a hurdle for designing a practical resource allocation algorithm, even for the case of offline resource allocation. In order to strike a balance between solution tractability and computational complexity,  we handle the above issues in two steps. First, we follow the approach in
\cite{JR:Roger_OFDMA} and relax $s_{i,k}(t)$ in constraint C7
to be a real value between zero and one instead of a Boolean, i.e., $0\le s_{i,k}(t)\le 1$. Then,
$s_{i,k}(t)$ can be interpreted as a time-sharing factor for
the $K$ users to utilize subcarrier $i$.   For facilitating the time sharing on each
subcarrier, we introduce three new variables and define them as
$\tilde{P}_{i,k}^E(t)={P}_{i,k}^E(t)s_{i,k}(t)$,
$\tilde{P}_{i,k}^N(t)={P}_{i,k}^N(t)s_{i,k}(t)$, and $\tilde{P}_{i,k}(t)={P}_{i,k}(t)s_{i,k}(t)$. These variables
represent the actual transmitted powers in the RF of the BS on subcarrier
$i$ for user $k$ under the time-sharing assumption.  Although the relaxation of the subcarrier allocation constraint
will generally result in a suboptimal solution, the authors in
\cite{JR:limited_backhaul,JR:Duality_gap}  show that the duality gap (sub-optimality) becomes zero when the number of subcarriers
is sufficiently large
for any multicarrier system that satisfies time-sharing\footnote{The proposed offline solution is asymptotically optimal when the number of subcarriers is large \cite{JR:limited_backhaul,JR:Duality_gap}. In fact, it has been shown in \cite{CN:large_subcarriers} via simulation that
 the duality gap is virtually zero for only 8 subcarriers in an
OFDMA system. Besides, the number of subcarriers employed in practical systems such as LTE  is in the order of hundreds. In other words, the solution obtained under the relaxed time-sharing  problem formulation is asymptotically optimal with respect to the original problem formulation.  On the other hand, we note that although time-sharing relaxation is assumed, the solution in (\ref{eqn:sub_selection}) indicates that the subcarrier allocation is still a Boolean which satisfies the binary constraint on the subcarrier allocation of the original problem.    }. Second, we introduce the following lemma which provides  valuable insight about the time varying dynamic of the optimal resource allocation policy.

\begin{Lem}\label{lem:optimal_hold} The optimal offline resource allocation policy\footnote{Here, ``optimality" refers to the optimality for the problem formulation under the time-sharing assumption.  } maximizing the system weighted energy efficiency does not change within an epoch.
\end{Lem}

 \emph{\quad Proof:}
Please refer to the Appendix  for a proof of Lemma \ref{lem:optimal_hold}.

As revealed by Lemma \ref{lem:optimal_hold}, the optimal resource allocation policy maximizing the weighted system energy efficiency is a constant in each epoch. Therefore, we can discretize the integrals and continuous variables in (\ref{eqn:inner_loop}). In other words, the number of constraints  in (\ref{eqn:inner_loop}) reduce to countable quantities. Without loss of generality, we assume that the channel states change $M\ge 0$ times and energy arrives $N\ge 0$ times in the duration of $[0, T]$.  Hence, we have $L=M+N$ epoch(s) for the considered duration of $T$ seconds. Time instant $T$ is treated as an extra  fading epoch with zero channel gains for all users to terminate the process. We define  the length of an epoch as $l_j = t_{j}-t_{j-1}$ where epoch $j\in\{1,2,\ldots,M+N\}$ is defined as the time interval $[t_{j-1},t_{j})$, cf. Figure \ref{fig:epoch}. Note that $t_{0}$ is defined as $t_{0}=0$. For the sake of notational simplicity and clarity, we replace all continuous-time variables with corresponding discrete time variables, i.e.,   $\tilde{P}_{i,k}(t) \rightarrow \tilde{P}_{i,k}[j]$, $\tilde{P}_{i,k}^E(t) \rightarrow \tilde{P}_{i,k}^E[j]$, $\tilde{P}_{i,k}^N(t) \rightarrow \tilde{P}_{i,k}^N[j]$, $P_C^E(t)\rightarrow P_C^E[j]$,  $P_C^N(t)\rightarrow P_C^N[j]$, $s_{i,k}(t)\rightarrow s_{i,k}[j]$, $H_{i,k}(t)\rightarrow H_{i,k}[j]$, $g_{k}(t)\rightarrow g_{k}[j]$, and $\Gamma_{i,k}(t)\rightarrow \Gamma_{i,k}[j]$. Then, the \emph{weighted total system capacity} and the \emph{weighted total energy consumption} can be re-written as
\begin{eqnarray}
U({\cal P},  {\cal S})\hspace*{-2mm}&=&\hspace*{-2mm}\sum_{k=1}^K
 \alpha_k\sum_{i=1}^{n_F}\sum_{j=1}^{L}   l_j C_{i,k}[j]\mbox{ and}\\
 U_{TP}({\cal P}, {\cal S})\hspace*{-2mm}&=&\hspace*{-2mm}\sum_{j=1}^{L} l_j\Big(\phi P_C^E[j]+P_C^N[j]\Big)+
 \sum_{k=1}^K \sum_{i=1}^{n_F}
\sum_{j=1}^L l_j \varepsilon \Big( \phi\tilde{P}_{i,k}^E[j]  + \tilde{P}_{i,k}^N[j]\Big) ,
\end{eqnarray}
respectively, where $C_{i,k}[j]=s_{i,k}[j]W\log_2\Big(1+\frac{\tilde{P}_{i,k}[j]\Gamma_{i,k}[j]}{s_{i,k}[j]}\Big)$ is the channel capacity between the BS and user $k$ on subcarrier $i$ in epoch $j$.  As a result, the optimization problem in (\ref{eqn:inner_loop}) is transformed into to the following convex optimization problem:
\begin{eqnarray}\label{eqn:cap_discrete}
&&\hspace*{50mm}\max_{{\cal P},  {\cal S}
}\,\,  U({\cal P},  {\cal S})-q  U_{TP}({\cal P}, {\cal S})\\
 &&\hspace*{-7mm}\mbox{C1:}\sum_{i=1}^{n_F}\sum_{k=1}^K\sum_{j=1}^e l_j \varepsilon \tilde{P}_{i,k}^E[j] +\sum_{j=1}^e P_C^E[j]l_j\le \sum_{j=1}^{e}E_{in}[j],\, \,\, \forall e\in\{1,2,\ldots,M+N\}, \notag\,\\
  &&\hspace*{-7mm}\mbox{C2:} \sum_{j=1}^{r}E_{in}[j]-\sum_{i=1}^{n_F}\sum_{k=1}^K\sum_{j=1}^{r-1} \varepsilon l_{j }\tilde{P}_{i,k}^E[j]-\sum_{j=1}^{r-1} l_j P_C^E[j]\le E_{\max}, \, \forall r\in\{2,\ldots, M+N+1\}, \notag
\\
&&\hspace*{-7mm}\mbox{C3:}\sum_{i=1}^{n_F}\sum_{k=1}^K l_e \varepsilon \tilde{P}_{i,k}^N[e]+l_e P_C^N[e]  \le P_N l_e  ,\,  \forall e, \hspace*{10mm}\mbox{C4: } l_e P_C^E[e]+l_e P_C^N[e] =l_e P_C ,\quad  \forall e,  \notag
\\
&&\hspace*{-7mm}\mbox{C5:}\sum_{k=1}^K
\sum_{i=1}^{n_F}\sum_{j=1}^{L}   l_j C_{i,k}[j] \ge R_{\min}  ,\hspace*{30mm} \mbox{C6:}\sum_{i=1}^{n_F}\sum_{k=1}^K l_e \tilde{P}_{i,k}[e]  \le l_e P_{\max}, \, \forall e,\notag\\
&&\hspace*{-7mm}\mbox{C7:}\notag\,\, 0 \le s_{i,k}[e]\le 1, \, \forall e,i,k,\, \mbox{C8:} \sum_{k=1}^K s_{i,k}[e]\le 1, \, \forall e,i, \,\,\,\mbox{C9:} P_{i,k}^N[e],P_{i,k}^E[e], {P}_{C}^N[e],{P}_{C}^E[e]\ge 0,\,\forall i,k,e,
\end{eqnarray}
where $E_{in}[j]$ in C1 is defined as the energy which arrives in epoch $j$. Hence, $E_{in}[j]=E_a$ for some $a$ if event $j$ is an energy arrival and $E_{in}[j]=0$ if event $j$ is a channel gain change, cf. Figure \ref{fig:epoch}.  The transformed problem in (\ref{eqn:cap_discrete}) is jointly concave w.r.t. all
optimization variables\footnote{We can follow a similar approach as in Appendix A to prove the concavity of the above problem for the  considered discrete time model.}, and under some mild conditions
\cite{book:convex}, solving the dual
problem is equivalent to solving the primal problem.

\begin{Remark}
 Mathematically,
$l_e$   on both sides of the (in)equalities  in C3, C4, and C6  in (\ref{eqn:cap_discrete})
  can be cancelled. Nevertheless, we do think that it is desirable to keep $l_e$ in these constraints  since they preserve the physical meaning of  C6; the energy consumption constraints in the system in time duration $l_e$.
\end{Remark}

\vspace*{-4mm}
\subsection{Dual Problem Formulation}
In this subsection, we solve transformed optimization problem (\ref{eqn:cap_discrete}). For this purpose, we first
need the Lagrangian function of the primal problem. Upon rearranging
terms, the Lagrangian can be written
as\begin{eqnarray}
 \hspace*{-8mm}\notag&& \hspace*{-6mm}{\cal L}(\boldmath\textbf{$\gamma$},\boldmath\textbf{$\beta$},\rho,  \boldmath\textbf{$\mu$},\boldmath\textbf{$\nu$},\boldmath\textbf{$\psi$}, \boldmath\textbf{$\eta$},{\cal P}, {\cal
 S})=\sum_{j=1}^{L} \sum_{k=1}^K
 \alpha_k\sum_{i=1}^{n_F}  l_j (w_k+\rho)C_{i,k}[j]-\rho R_{\min}+\sum_{j=1}^L \sum_{i=1}^{n_F}\sum_{k=1}^K\eta_{i,j}\notag\\ \notag
&&\hspace*{-8mm}-\sum_{j=1}^L\gamma_j\Big(\sum_{i=1}^{n_F}\sum_{k=1}^K\sum_{m=1}^j \varepsilon l_m  \tilde{P}_{i,k}^E[m]+ \sum_{m=1}^j  l_m P_C^E[m] - \sum_{m=1}^{j}E_{in}[m]\Big)-\sum_{j=1}^L \nu_j l_j( P_C^E[j]+ P_C^N[j])\end{eqnarray}\begin{eqnarray}
\notag&& \hspace*{-8mm}-q\Big(\sum_{j=1}^{L} l_j\Big(\phi P_C^E[j]+P_C^N[j]\Big)+
 \sum_{k=1}^K \sum_{i=1}^{n_F}
\sum_{j=1}^L l_j \varepsilon \Big( \phi\tilde{P}_{i,k}^E[j]  + \tilde{P}_{i,k}^N[j]\Big)\Big)  +\sum_{j=1}^L \nu_j l_j P_C\notag \\
  &&\hspace*{-8mm}-\sum_{j=2}^{L+1}\beta_j\Big(\sum_{m=1}^{j}E_{in}[m]-\sum_{i=1}^{n_F}\sum_{k=1}^K\sum_{m=1}^{j-1} \varepsilon l_{m }\tilde{P}_{i,k}^E[m]-\sum_{m=1}^{j-1} l_m P_C^E[m]- E_{\max}\Big)-\sum_{j=1}^L \sum_{i=1}^{n_F}\sum_{k=1}^K\eta_{i,j}s_{i,k}[j]\notag
  \\
&&\hspace*{-8mm}-\sum_{j=1}^L\mu_j\Big(\sum_{i=1}^{n_F}\sum_{k=1}^K  \varepsilon l_j \tilde{P}_{i,k}^N[j]  + l_j P_C^N[j]  -l_j P_N \Big)-\sum_{j=1}^L \psi_j\Big(\sum_{i=1}^{n_F}\sum_{k=1}^K l_j \tilde{P}_{i,k}[j]  -l_j P_{\max} \Big),
 \label{eqn:Lagrangian}
\end{eqnarray}
where $\boldmath\textbf{$\gamma$}$ is the Lagrange multiplier vector associated with causality constraint C1 on consuming energy from the energy harvester and has elements $\gamma_j$, $j\in\{1,\ldots,L\}$. $\boldmath\textbf{$\beta$}$ is the Lagrange multiplier vector corresponding to the maximum energy level constraint C2 in the battery of the energy harvester with elements $\beta_{j}$ where $\beta_1=0$. $\rho$ is the Lagrange multiplier corresponding to the minimum data rate requirement $R_{\min}$ in C5. $\boldmath\textbf{$\mu$}$, $\boldmath\textbf{$\nu$}$,  and $\boldmath\textbf{$\psi$}$ have elements $\mu_{j}$, $\nu_{j}$, and $\psi_{j}$ are the Lagrange multiplier vectors for constraints C3, C4, and C6, respectively.  $\boldmath\textbf{$\eta$}$ is the Lagrange multiplier vector accounting for subcarrier usage
constraint C8 with elements $\eta_{i,j},i\in\{1,\ldots,n_F\}$. Note that the boundary constraints C7 and C9 are absorbed
into the Karush-Kuhn-Tucker (KKT) conditions when deriving the optimal solution in Section \ref{sect:sub_problem_solution}.

Thus, the dual problem is given by
\begin{eqnarray}\label{eqn:dual_problem}
\underset{\boldmath\textbf{$\gamma$},\boldmath\textbf{$\beta$},\rho,  \boldmath\textbf{$\mu$},\boldmath\textbf{$\psi$}, \boldmath\textbf{$\eta$} \ge
0}{\min}\ \underset{{\cal P},{\cal S}}{\max}\,\,
{\cal L}(\boldmath\textbf{$\gamma$},\boldmath\textbf{$\beta$},\rho,  \boldmath\textbf{$\mu$},\boldmath\textbf{$\nu$}, \boldmath\textbf{$\psi$}, \boldmath\textbf{$\eta$},{\cal P}, {\cal
 S}).\label{eqn:master_problem}
\end{eqnarray}
Note that $\boldmath\textbf{$\nu$}$ is not an optimization variable in (\ref{eqn:dual_problem}) since C4 in (\ref{eqn:cap_discrete}) is an equality constraint.

\subsection{Dual Decomposition and Subproblem Solution}
\label{sect:sub_problem_solution} By Lagrange dual decomposition, the dual
problem is decomposed into two parts (nested loops): the
first part (inner loop) consists of $n_F+1$ subproblems where $n_F$ subproblems have
identical structure; the second part (outer loop) is the master dual
problem. The dual problem can be
solved iteratively where in each iteration the BS solves $ n_F$
 subproblems (inner loop) in parallel and solves the master problem (outer loop) with the gradient method.

Each one of the $n_F$  subproblems with identical structure is designed for one subcarrier and can be
expressed as
\begin{eqnarray}
\label{eqn:sub-problem} &\underset{{\cal P}, { \cal
S}}{\max}&{\cal L}_i(\boldmath\textbf{$\gamma$},\boldmath\textbf{$\beta$},\rho,  \boldmath\textbf{$\mu$},\boldmath\textbf{$\nu$},\boldmath\textbf{$\psi$}, \boldmath\textbf{$\eta$},{\cal P}, {\cal
 S})\end{eqnarray}
for a fixed set of Lagrange multipliers
where
\begin{eqnarray}
\notag &&\hspace*{-4mm}{\cal {L}}_{i}({\boldmath\textbf{$\gamma$}},
{\boldmath\textbf{$\beta$}},\rho,
{\boldmath\textbf{$\mu$}},{\boldmath\textbf{$\nu$}}, {\boldmath\textbf{$\psi$}},
{\boldmath\textbf{$\eta$}}, {\cal P}, {\cal S})=\sum_{j=1}^{L} \sum_{k=1}^K
 \alpha_k  l_j (w_k+\rho)C_{i,k}[j]\hspace*{-1mm}+\hspace*{-1mm}\sum_{j=2}^{L+1}\beta_j\sum_{m=1}^{j-1} l_m P_C^E[m]\notag\\
  &&\hspace*{-5mm}+\sum_{j=2}^{L+1}\beta_j\sum_{k=1}^K\sum_{m=1}^{j-1} \varepsilon l_{m }\tilde{P}_{i,k}^E[m]-q\Big(
 \sum_{k=1}^K
\sum_{j=1}^L l_j\varepsilon \Big( \phi \tilde{P}_{i,k}^E[j]  +\tilde{P}_{i,k}^N[j]\Big)+\sum_{j=1}^{L}l_j \Big(\phi P_C^E[j]+P_C^N[j]\Big)\Big)\notag\end{eqnarray}\begin{eqnarray}
&&\hspace*{-3cm}-\hspace*{-0.5mm}\sum_{j=1}^L \psi_j\Big(\sum_{k=1}^K l_j\tilde{P}_{i,k}[j]  \Big)
-\hspace*{-0.5mm}\sum_{j=1}^L\eta_{i,j} \Big(\sum_{k=1}^K s_{i,k}[j]\Big)-\sum_{j=1}^L\mu_j\Big(\sum_{k=1}^K  \varepsilon l_j \tilde{P}_{i,k}^N[j]  +l_j P_C^N[j] \Big)\notag\\
&&\hspace*{-3cm}-\sum_{j=1}^L l_j \nu_j(P_C^E[j]+ P_C^N[j])-\sum_{j=1}^L\gamma_j\Big(\sum_{k=1}^K\sum_{m=1}^j \varepsilon l_m  \tilde{P}_{i,k}^E[m]+\sum_{m=1}^j l_m P_C^E[m]\Big).
\end{eqnarray}

Let $\tilde{P}_{i,k}^{E*}[j],\tilde{P}_{i,k}^{N*}[j],{P}_{C}^{E*}[j],{P}_{C}^{N*}[j]$, and $s_{i,k}^*[j]$ denote the solution of  subproblem (\ref{eqn:sub-problem}) for event $j$. Using standard optimization
techniques and the KKT conditions, the power allocation
 for  signal transmission for
user $k$ on subcarrier $i$ for event $j$  is given by
\begin{eqnarray}\label{eqn:Final_power_allocation1}\notag
\tilde{P}_{i,k}^{E*}[j]\hspace{-2mm}&=&\hspace{-2mm} s_{i,k}[j]{P}_{i,k}^{E*}[j]= s_{i,k}[j]\Bigg[\frac{W(\alpha_k+\rho)}{\ln(2)(\sum_{e=j}^L\gamma_e\varepsilon-\sum_{e=j}^{L}\beta_{e+1}\varepsilon+q \phi\varepsilon+\psi_j)}
-\frac{1}{\Gamma_{i,k}[j]}\Bigg]^+\,\mbox{and}\,\,\\
 \tilde{P}_{i,k}^{N*}[j]\hspace{-2mm}&=&\hspace{-2mm} s_{i,k}[j]{P}_{i,k}^{N*}[j]= s_{i,k}[j]\Bigg[\frac{W(\alpha_k+\rho)}{\ln(2)(q\varepsilon+\mu_j\varepsilon+\psi_j)}
-\frac{1}{\Gamma_{i,k}[j]}-\tilde{P}_{i,k}^{E*}[j]\Bigg]^+, \label{eqn:Final_power_allocation2}
 \end{eqnarray}
for $\phi< 1$. The power allocation solution  in (\ref{eqn:Final_power_allocation1}) can be interpreted as a \emph{multi-level} water-filling scheme as the water levels of different users can be different.
Interestingly, the  value of $\tilde{P}_{i,k}^{N*}[j]$ depends on $\tilde{P}_{i,k}^{E*}[j]$. As can be seen in (\ref{eqn:Final_power_allocation2}), $\tilde{P}_{i,k}^{E*}[j]$ decreases the water-level for calculation of the value of $\tilde{P}_{i,k}^{N*}[j]$. In other words,  $\tilde{P}_{i,k}^{E*}[j]$ reduces the amount of energy drawn from the non-renewable source for maximization of energy efficiency. Besides, it can be observed from (\ref{eqn:Final_power_allocation1}) that the BS does not always consume all available renewable energy in each epoch for maximization of the weighted energy efficiency  and the value of $q$ determines at what point the water-level is clipped. On the other hand, in order to obtain the subcarrier allocation, we
take the derivative of the subproblem w.r.t. $s_{i,k}[j]$, which yields $\frac{\partial  {\cal {L}}_{i}(\ldots)}{\partial
s_{i,k}^*[j]}=Q_{i,k}[j]-\eta_{i,j}$, where $Q_{i,k}[j]\ge0$ is the marginal benefit
\cite{JR:Marginal_benefit} for allocating subcarrier $i$ to user $k$ for event $j$
and is given by $Q_{i,k}[j]=$
\begin{eqnarray} \label{eqn:subcarrier_allocation}
W(\alpha_k+\rho)\Bigg(\log_2\Big(1+\Gamma_{i,k}[j]({P}_{i,k}^{E*}[j]+{P}_{i,k}^{N*}[j])\Big)-\frac{\Gamma_{i,k}[j]({P}_{i,k}^{E*}[j]+{P}_{i,k}^{N*}[j])}{\ln(2)(1+\Gamma_{i,k}[j]({P}_{i,k}^{E*}[j]+{P}_{i,k}^{N*}[j]))}\Bigg).
\end{eqnarray}  Thus, the subcarrier selection
 on subcarrier $i$ in event $j$ is given by
\begin{eqnarray}\hspace*{-1mm}
\label{eqn:sub_selection}s_{i,k}^*[j]=
 \left\{ \begin{array}{rl}
 1 &\mbox{if $k=\arg \underset{c}{\max} \,\, \ Q_{i,c}[j]$ }  \\
 0 &\mbox{ otherwise}
       \end{array} \right.\hspace*{-3mm} .
\end{eqnarray}
It can be observed from (\ref{eqn:subcarrier_allocation}) that only the user who can provide the largest marginal benefit on subcarrier $i$ in epoch $j$ is selected by the resource allocator, for transmission on that subcarrier.  This is because the channel gains of different users are generally different due to uncorrelated fading across different users. We note that a larger marginal benefit is not necessarily equivalent to a larger system
 throughput since the marginal benefit includes a notion of fairness.

 After solving the $n_F$ subproblems with identical structure, we calculate the amount of power used for signal processing in each of the two energy sources.
We substitute $P_C^{N}[j]=P_C-P_C^{E}[j]$ into       (\ref{eqn:sub-problem}) which yields the following KKT condition for $P_C^{E*}[j]$:
        \begin{eqnarray}\label{eqn:power_signal_processing_KKT}
       \frac{\partial {\cal L}_i(\ldots)}{\partial {P}_{C}^{E*}[j]}=-l_j\sum_{e=j}^L\gamma_e+l_j\sum_{e=j}^{L}\beta_{e+1}  -q l_j\phi +q l_j +\mu l_j\left\{ \begin{array}{rl}
 \ge 0, &{P}_{C}^{E*}[j]\ge 0 \\
 <0, &\mbox{ otherwise}
       \end{array} \right..
 \end{eqnarray}
 It can be observed from (\ref{eqn:power_signal_processing_KKT}) that the Lagrangian function ${\cal L}_i(\ldots)$ is an affine function in  $P_C^{E*}[j]$. In other words, the value of $P_C^{E*}[j]$ must be one of the two vertexes of a feasible solution set created by the associated constraints.
  As a result, the powers used for signal processing drawn from the energy harvester and the non-renewable source are given by
\begin{eqnarray}\label{eqn:Final_power_circuit_1}
{P}_{C}^{E*}[j] &=&\Bigg[\frac{\sum_{a=1}^{j}E_{in}[a]-\sum_{i=1}^{n_F}\sum_{k=1}^K\sum_{a=1}^j l_a \varepsilon \tilde{P}_{i,k}^{E*}[a]- \sum_{m=1}^{j-1} P_C^E[m]l_m}{l_j}\Bigg]_0^{P_C} \,\mbox{and}\,\,\\
 {P}_{C}^{N*}[j]&=&P_C-{P}_{C}^{E*}[j],\label{eqn:Final_power_circuit_2}
 \end{eqnarray}
respectively. The numerator of variable ${P}_{C}^{E*}[j]$ in (\ref{eqn:Final_power_circuit_1}) represents the residual energy level in the battery, i.e., the vertexes (feasible set) created by the associated constraints on ${P}_{C}^{E*}[j]$. Equations  (\ref{eqn:Final_power_circuit_1}) and (\ref{eqn:Final_power_circuit_2}) indicate that if the amount of energy in the energy harvester is not sufficient to fully supply the required energy  $P_C$, i.e, ${P}_{C}^{E*}[j]<P_C$, then the BS will also draw energy from the non-renewable energy source such that ${P}_{C}^{E*}[j]+{P}_{C}^{N*}[j]=P_C$.

\subsection{Solution of the Master Dual Problem}
 For solving the master minimization problem
in (\ref{eqn:master_problem}), i.e, to find
{\boldmath\textbf{$\gamma$}}, {\boldmath\textbf{$\beta$}}, $\rho$, {\boldmath\textbf{$\mu$}}, and {\boldmath\textbf{$\psi$}} for given ${\cal P}$ and ${\cal  S}$, the gradient method can be used since
the dual function is differentiable. The gradient update equations
are given by:
\begin{eqnarray}\label{eqn:multipler1}
\hspace*{-5mm}\gamma_j(\varsigma+1)\hspace*{-3mm}&=&\hspace*{-3mm}\Big[\gamma_j(\varsigma)-\xi_1(\varsigma)\hspace*{-1mm} \times  \hspace*{-1mm}
\Big( \sum_{m=1}^{j}E_{in}[m] -\sum_{m=1}^j P_C^E[m]l_m\sum_{i=1}^{n_F}\sum_{k=1}^K\sum_{m=1}^j l_m \varepsilon \tilde{P}_{i,k}^E[m] \Big)\Big]^+,\forall j,\\
\label{eqn:multipler2}
\hspace*{-5mm} \beta_r(\varsigma+1)\hspace*{-3mm}&=&\hspace*{-3mm}\Big[\beta_r(\varsigma)-\xi_2(\varsigma)\hspace*{-1mm} \times  \hspace*{-1mm}
\Big(\hspace*{-1mm}E_{\max}\hspace*{-1mm}+\hspace*{-1mm}\sum_{m=1}^{r-1}P_C^E[m]l_m \hspace*{-1mm}-\hspace*{-1mm}\sum_{m=1}^r E_{in}[m]\hspace*{-1mm}+\hspace*{-1mm}\sum_{i=1}^{n_F}\sum_{k=1}^K\sum_{m=1}^r\hspace*{-1mm} \varepsilon l_{m }\tilde{P}_{i,k}^E[m]\Big)\hspace*{-1mm}\Big]^+\hspace*{-1mm},\hspace*{-1mm}\forall r,\\
\hspace*{-5mm}\rho(\varsigma+1)\hspace*{-3mm}&=&\hspace*{-3mm}\Big[\rho(\varsigma)-\xi_3(\varsigma)\hspace*{-1mm} \times  \hspace*{-1mm}
\Big(\sum_{j=1}^{L} \sum_{k=1}^K
\sum_{i=1}^{n_F}  l_j C_{i,k}[j]- R_{\min}\Big)\Big]^+,
\label{eqn:multipler3}\\
\hspace*{-5mm}\mu_j(\varsigma+1)\hspace*{-3mm}&=&\hspace*{-3mm}\Big[\mu_j(\varsigma)-\xi_4(\varsigma)\hspace*{-1mm} \times  \hspace*{-1mm}
\Big(P_N l_j-\sum_{i=1}^{n_F}\sum_{k=1}^K  \varepsilon \tilde{P}_{i,k}^N[j]  l_j-P_C^N[j]l_j \Big)\Big]^+, \forall j,
\label{eqn:multipler4}\\
\hspace*{-50mm}\psi_j(\varsigma+1)\hspace*{-3mm}&=&\hspace*{-3mm}\Big[\psi_j(\varsigma)-\xi_5(\varsigma)\hspace*{-1mm} \times  \hspace*{-1mm}
\Big(P_{\max}l_j - \sum_{i=1}^{n_F}\sum_{k=1}^K \tilde{P}_{i,k}[j]l_j \Big)\Big]^+, \forall j,
\label{eqn:multipler6}
\end{eqnarray}
where $j\in\{1,\ldots\,M+N\}$ and $r\in\{2,\ldots\,M+N\}$. $\varsigma\ge 0$ and $\xi_u(\varsigma)$,
$u\in\{1,\ldots,5\}$, are the iteration index and  positive step sizes, respectively.  The updated Lagrange
multipliers in (\ref{eqn:multipler1})-(\ref{eqn:multipler6}) are
used for solving the subproblems in
(\ref{eqn:master_problem}) via updating the resource allocation
policies. Updating $\eta_{i,j}$ is not necessary as it has
the same value for all users and does not affect the subcarrier
allocation in (\ref{eqn:sub_selection}).  On the other hand,  since the transformed problem in (\ref{eqn:cap_discrete}) is jointly concave w.r.t. the optimization variables and satisfies Slater's constraint qualification \cite{book:convex},  the duality gap between dual optimal and relaxed primal optimal is zero and it is guaranteed
that the iteration between the master problem and the subproblems converges to the
 solution of (\ref{eqn:inner_loop}) in the main loop, if
the chosen step
 sizes satisfy the infinite travel condition
 \cite{book:convex,Notes:Sub_gradient}.

 Note that although the proposed asymptotically (i.e., for a sufficiently large number of subcarriers) optimal offline algorithm requires non-causal knowledge of the channel gains and energy arrivals which may not be available in practice, the performance of the asymptotically optimal offline algorithm serves as an upper bound for any online scheme. Besides, the structure of the asymptotically optimal offline algorithm sheds some light on the design of online algorithms. In the next section, we will address the causality issue by studying two online  resource allocation algorithms which utilize causal energy arrival and channel gain information only.

%
\vspace*{-2mm}
\section{ Online Resource Allocation and Scheduling Design}
\label{sect:4}

In this section, we study the  optimal and a suboptimal online resource allocation algorithms requiring only causal  information of energy arrivals and channel states.
\subsection{Optimal Online Solution}\label{sect:DP}
In practice, the instantaneous  CSI of the users is available at the BS, and can be obtained via feedback  and exploiting channel reciprocity in frequency division duplex (FDD) systems and time division duplex (TDD) systems, respectively. Besides, the current energy arrival information for each \emph{energy event} is available after the energy has been harvested. On the contrary, the future CSI and future energy arrival information are not available when the BS computes the resource allocation policy.  Therefore, we adopt a statistical approach in the following problem formulation. The optimal online resource allocation policy $\{\cal P, S\}$ can be obtained by
 maximizing the \emph{expected weighted energy efficiency}:
\begin{eqnarray}\label{eqn:online-formulation1}
&& \notag \hspace*{45mm}
\max_{{\cal P}, {\cal S}}\,{\cal E}_{\mathbf{F,E}}\Big\{U_{eff}({\cal P}, {\cal S}) \Big\} \\
\notag \mbox{s.t.} &&\hspace*{38mm}\quad\mbox{C3,\,C4,\,C6,\,C7,\,C8,\,C9} \nonumber\\
&&\hspace*{-8mm}\notag\mbox{C1:} {\cal E}_{\mathbf{F,E}}\Big\{\sum_{i=1}^{n_F}\sum_{k=1}^K\int_0^{t^E_b -\delta} \varepsilon  s_{i,k}(u)P_{i,k}^E(u)\, du+\int_0^{t^E_b -\delta} P_C^E(u)\, du  \Big\} \hspace*{-1mm}\le \hspace*{-1mm}\sum_{j=0}^{b-1}E_j,\, \forall b\in\{1,2,\ldots\},\\
&&\hspace*{-8mm}\notag\mbox{C2:} {\cal E}_{\mathbf{F,E}}\Big\{\sum_{j=0}^{d(t)}E_j-\sum_{i=1}^{n_F}\sum_{k=1}^K\int_0^{t} \varepsilon  s_{i,k}(u)P_{i,k}^E(u)\, du-\int_0^t P_C^E(u)\, du\Big\}\le E_{\max}, \,  0\le t\le T,\\
&&\hspace*{-8mm}\mbox{C5:}\,{\cal E}_{\mathbf{F,E}}\Big\{\sum_{k=1}^K
\sum_{i=1}^{n_F}
C_{i,k}\Big\} \ge R_{\min},
\end{eqnarray}
where vectors $\mathbf{E}$ and $\mathbf{F}$  in (\ref{eqn:online-formulation1}) contain the random  energy arrivals and channel gains, respectively.  Note that although constraints C3, C4, and C6--C9 are the same as in the case of offline algorithm design in (\ref{eqn:cross-layer}), the problem formulation here is different from (\ref{eqn:cross-layer}). First, C5 specifies now the minimum required \emph{average} data rate of the system. Besides, constraints C1 and C2 are imposed to constrain the \emph{average} energy usage of the system, instead of the instantaneous energy consumption.

 For solving (\ref{eqn:online-formulation1}), we first   define $\{\cal \bar{P}, \bar{S}\}$ as a feasible resource allocation policy  which satisfies constraints C1--C9  in (\ref{eqn:online-formulation1}). Also,  we denote the amount of energy available in the battery at time $t$ by $e(t)$.  Then,
we apply Theorem 1 to transform the objective function from fractional form into subtractive form.  The resulting objective function can be written as
 \begin{eqnarray}\notag
J({\cal \bar{P},\bar{S}},t,e(t))= &&\hspace*{-2mm}{\cal E}_{\mathbf{F,E}}\Bigg\{\sum_{k=1}^K
 \alpha_k\sum_{i=1}^{n_F}W\int_{t}^T s_{i,k}(\tau) C_{i,k}(\tau) d\tau -q \Big[\Big.\Big. \int_t^T \Big(  \phi P_C^E(\tau)+P_C^N(\tau)\Big)\, d\tau\\
 &&\hspace*{-2mm}\Big. +
 \sum_{k=1}^K \sum_{i=1}^{n_F}
  \int_{t}^T s_{i,k}(\tau)\varepsilon\Big( P_{i,k}^E(\tau) \phi+ P_{i,k}^N(\tau)\Big)\, d\tau\Big]\Bigg\},
 \end{eqnarray}
 where $t=0$, $C_{i,k}(\tau)=\log_2\Big(1+P_{i,k}(\tau)\Gamma_{i,k} (\tau)\Big)$, and $q$ can be found via a similar approach as described in Table I. After that, we can  approximate
the integrals in (\ref{eqn:online-formulation1}) as  Riemann sums of  $\Xi$ equally spaced intervals with an interval width of $\epsilon=\frac{T}{\Xi}$. Thus, for a sufficiently small value of
 $\epsilon$, we can discretize the integrals and the objective function in (\ref{eqn:online-formulation1}) can be parameterized by $t$ \cite{JR:DP_discrete}:
  \begin{eqnarray}\notag
&&J({\cal \bar{P},\bar{S}},m \epsilon,e(m \epsilon))= {\cal E}_{\mathbf{F,E}}\Bigg\{\sum_{k=1}^K
 \alpha_k\sum_{i=1}^{n_F}W \sum_{\upsilon
=m}^{\Xi-1 }\epsilon s_{i,k}(\upsilon
\epsilon ) C_{i,k}(\upsilon
\epsilon) \\
 &&\hspace*{-2mm} -q\epsilon \Big[\Big.\Big. \sum_{\upsilon
=m}^{\Xi-1 } \Big(  \phi P_C^E(\upsilon
\epsilon)+P_C^N(\upsilon
\epsilon)\Big)\,\Big. +
 \sum_{k=1}^K \sum_{i=1}^{n_F}
 \sum_{\upsilon
=m}^{\Xi-1 } \epsilon s_{i,k}(\upsilon
\epsilon)\varepsilon\Big( P_{i,k}^E(\upsilon
\epsilon) \phi+ P_{i,k}^N(\upsilon
\epsilon)\Big)\,\Big]\Bigg\},
 \end{eqnarray}
where $t=m \epsilon $ for $m=\{0,1,2,\ldots,\Xi-1\}$. Then, the  optimal  \emph{cost-to-go} function  \cite{book:DP_vol1,JR:constrainted_DP} at time  $t$ is given by
 \begin{eqnarray}
 J^*(m \epsilon,e(m \epsilon))= \underset{{\cal \bar{P},\bar{S}}}{\max}\,\,  J({\cal \bar{P},\bar{S}},m \epsilon,e(m \epsilon)).
 \end{eqnarray}
By applying Bellman's
equations and backward induction \cite{book:DP_vol1,JR:constrainted_DP}, it can be shown that the optimal resource allocation policy $\{\bar{\cal P}^*,\bar{\cal S}^*\}$ for solving $(\ref{eqn:online-formulation1})$ must satisfy the following dynamic programming (DP) equation:
 \begin{eqnarray}\label{eqn:dp}
&&\hspace{-2mm} \, \,J^*(m \epsilon,e(m \epsilon))\\ \notag
\hspace{-2mm}&=& \hspace{-2mm}\,\underset{{\cal \bar{P},\bar{S}}}{\max} \,\Bigg\{\epsilon\sum_{k=1}^K
 \alpha_k\sum_{i=1}^{n_F}W s_{i,k}(m \epsilon) C_{i,k}(m \epsilon)-q\epsilon\Big[ \Big(  \phi P_C^E(m \epsilon)+P_C^N(m \epsilon)\Big)\Big]\Bigg.\\
  &&\hspace{-2mm}-q\epsilon\Big[
 \sum_{k=1}^K \sum_{i=1}^{n_F}
  s_{i,k}(m \epsilon)\varepsilon\Big( P_{i,k}^E(m \epsilon) \phi+ P_{i,k}^N(m \epsilon)\Big)\Big]+J^*((m+1) \epsilon,e((m+1) \epsilon))\Bigg\},\quad \forall m.\notag
 \end{eqnarray}
 In other words, the optimal online resource allocation policy in (\ref{eqn:online-formulation1}) can be obtained by solving  (\ref{eqn:dp}) via standard DP \cite{book:DP_vol1,JR:constrainted_DP}. To summarize the implementation of the optimal online resource allocation algorithm, at time $t=0$, the BS computes the
resource allocation policy via standard DP. Note that the  policy is a function of CSI $H_{i,k}(t)$ and the energy level $e(t)$ in the  battery. For time $t>0$, the BS updates  $H_{i,k}(t)$ and $e(t)$ and performs resource allocation based on $\{\bar{\cal P}^*,\bar{\cal S}^*\}$ at each discretized time interval $t=m \epsilon $ for each value of $m$.

\vspace*{-3mm}
\subsection{Suboptimal Online Solution}
In the last section, we introduced the optimal online resource allocation policy which can be obtained via DP. Yet, it is well known that DP suffers from the ``\emph{curse of dimensionality}". Specifically, the search space for the optimal solution increases
exponentially  w.r.t. the number of users and subcarriers. Hence,
in practice, DP is not applicable for the considered system due to the huge computational complexity and memory
requirement. In the following, we propose a suboptimal online resource allocation algorithm which is inspired by the asymptotically optimal offline resource allocation derived in Section \ref{sect:solution_dual_decomposition}. In particular, the proposed suboptimal resource allocation algorithm is \emph{event-driven} and each computation is triggered by a change in fading level or an energy arrival.
In other words, the proposed suboptimal online algorithm requires only  causal system information  and the statistics of the involved events which leads to a lower complexity compared to the optimal online solution. Note that  in order to emphasize  the similarity between the offline algorithm and the proposed suboptimal online algorithm, with a slight abuse of notation, we use a similar notation for both algorithms.

\subsubsection*{Suboptimal Algorithm}
The structure of the  offline resource allocation algorithm in Section \ref{sect:3} depends on the length of each epoch. However, this knowledge is unavailable at the BS  due to causality constraints. As a compromise solution, we focus on the statistical average of the length of each event. We define the average length of each epoch as $\overline{L_E}$ and there are $Z=\lfloor\frac{T}{\overline{L_E}}+1\rfloor$ events in $T$ seconds on average.
In practice, the value of $\overline{L_E}$  can be estimated by long term channel and energy arrival measurements. Besides, to simplify the resource allocation algorithm, we assume that the resource allocation policy is constant in each epoch which was shown to be optimal for the case of offline algorithm design. As a result, we can directly formulate the resource allocation design  problem by using a discrete representation.
Without loss of generality, we focus on the resource allocation algorithm design for epoch $j$. Then, the \emph{weighted average system throughput} and the \emph{total weighted energy consumption} in epoch $j$ are given by
\begin{eqnarray}
 U({\cal P}_j,  {\cal S}_j)\hspace*{-2mm}&=&\hspace*{-2mm}\sum_{k=1}^K
 \alpha_k\sum_{i=1}^{n_F}\overline{L_E} C_{i,k}[j]\mbox{ and}\notag\\
 U_{TP}({\cal P}_j, {\cal S}_j)\hspace*{-2mm}&=&\hspace*{-2mm}\overline{L_E}\Big(\phi P_C^E[j]+P_C^N[j]\Big)+
 \sum_{k=1}^K \sum_{i=1}^{n_F}
\overline{L_E} \varepsilon \Big( \tilde{P}_{i,k}^E[j] \phi + \tilde{P}_{i,k}^N[j]\Big) ,
\end{eqnarray}
respectively.
The  resource allocation policy, ${\cal P}_j=\{ \tilde{P}_{i,k}^E[j], \tilde{P}_{i,k}^N[j], P_C^E[j], P_C^N[j]\}$, ${\cal S}_j=\{ s_{i,k}[j]\}$, which maximizes the weighted energy efficiency in epoch $j$ can be
obtained by solving
\begin{eqnarray}
\label{eqn:cross-layer-offline}&&\hspace*{25mm} \max_{{\cal P}_j,  {\cal S}_j
}\,\, \frac{ U({\cal P}_j,  {\cal S}_j)}{ U_{TP}({\cal P}_j, {\cal S}_j)}\\
\notag \hspace*{20mm}\mbox{s.t.} &&\hspace*{10mm}\mbox{C1:}\sum_{i=1}^{n_F}\sum_{k=1}^K \overline{L_E}\varepsilon \tilde{P}_{i,k}^E[j]+\overline{L_E} P_C^E[j]  \le E[j],\notag\\  &&\hspace*{-33mm}\mbox{C3:}\sum_{i=1}^{n_F}\sum_{k=1}^K \overline{L_E}\varepsilon \tilde{P}_{i,k}^N[j]+\overline{L_E} P_C^N[j]  \le P_N \overline{L_E},  \hspace*{12mm}\mbox{C4: }  \overline{L_E} P_C^E[j]+ \overline{L_E}P_C^N[j] =\overline{L_E} P_C,  \notag
\\
&&\hspace*{-33mm}\mbox{C5:}\sum_{k=1}^K
\sum_{i=1}^{n_F}\sum_{j=1}^{L}  \overline{L_E} C_{i,k}[j] \ge \frac{R_{\min}}{Z}  ,\hspace*{29mm} \mbox{C6:}\sum_{i=1}^{n_F}\sum_{k=1}^K \overline{L_E} \tilde{P}_{i,k}[j]  \le \overline{L_E} P_{\max},\notag\\
&&\hspace*{-33mm}\mbox{C7:}\notag\,\, 0 \le s_{i,k}[j]\le 1, \, \forall i,k,\hspace*{3mm} \mbox{C8:} \sum_{k=1}^K s_{i,k}[j]\le 1, \, \forall i, \,\,\,\mbox{C9:} P_{i,k}^N[j],P_{i,k}^E[j], {P}_{C}^N[j],{P}_{C}^E[j]\ge 0,\,\forall i,k,
\end{eqnarray}
where $E[j]$ is the amount of energy available in the battery in epoch $j$.  It captures the joint effect of channel fluctuations, energy arrivals, and resource allocation  in the previous epochs on the energy availability in epoch $j$. This information is available at the BS by monitoring the amount of energy consumed and harvested in the past epochs. Note that the battery overflow constraint C2 is not imposed in the suboptimal online problem  formulation (\ref{eqn:cross-layer-offline}) for solution tractability. In practice, the amount of energy exceeding the battery storage will be discharged and not stored. The performance loss caused by  the above problem formulation compared to the optimal one will be investigated  in the simulation section.

To solve the optimization problem in (\ref{eqn:cross-layer-offline}),
we can use Theorem 1 (objective function transformation) and Algorithm 1 (iterative algorithm) which were introduced in Section \ref{sect:algorithm}. In particular,
 in each iteration of the main loop, cf. Table I, we solve
the following optimization problem for a given parameter $q$:
\begin{eqnarray}\label{eqn:lag_offline}
&&\hspace*{-20mm}\max_{{\cal P}_j, {\cal S}_j}
\quad\,{U}({\cal P}_j, {\cal S}_j)-q{U}_{TP}({\cal P}_j, {\cal
 S}_j)\nonumber\\
&&\mbox{s.t.} \,\,\mbox{C1, C3--C9}.
\end{eqnarray}
The above
optimization problem can be proved to be jointly concave w.r.t. the optimization variables by using a similar approach as in  the Appendix. Similar to the offline resource allocation problem, we solve (\ref{eqn:lag_offline}) by dual decomposition. The Lagrangian of (\ref{eqn:lag_offline}) is given by
\begin{eqnarray}\notag
&&\hspace*{-5mm}{\cal L}(\gamma,\psi,\rho,  \boldmath\textbf{$\eta$},\mu,\nu,{\cal P}_j, {\cal
 S}_j)= \sum_{k=1}^K
 \alpha_k\sum_{i=1}^{n_F}  \overline{L_E} (w_k+\rho) C_{i,k}[j]-\rho \frac{R_{\min}}{Z}-q\overline{L_E}\Big(\Big(\phi P_C^E[j]+P_C^N[j]\Big) \\ \notag
 &&\hspace*{-5mm}+
 \sum_{k=1}^K \sum_{i=1}^{n_F}
 \varepsilon \Big( \tilde{P}_{i,k}^E[j] \phi + \tilde{P}_{i,k}^N[j]\Big)\Big)-\gamma\Big(\sum_{i=1}^{n_F}\sum_{k=1}^K \overline{L_E}\varepsilon \tilde{P}_{i,k}^E[j]+\overline{L_E} P_C^E[j] - E[j]\Big)\\
&& \hspace*{-5mm}-\mu\overline{L_E}\Big(\sum_{i=1}^{n_F}\sum_{k=1}^K \varepsilon \tilde{P}_{i,k}^N[j]+ P_C^N[j]- P_N \Big) -\nu\overline{L_E}\Big(P_C^E[j]+P_C^N[j]-P_C\Big)  \notag
\\
&&\hspace*{-5mm}-\psi \overline{L_E}\Big(\sum_{i=1}^{n_F}\sum_{k=1}^K \tilde{P}_{i,k}[j]  -  P_{\max}\Big)-\sum_{i=1}^{n_F} \eta_i\Big( \sum_{k=1}^K s_{i,k}[j]-1\Big),
 \label{eqn:Lagrangian-online}
\end{eqnarray}
where $\gamma,\,\mu,\,\nu,\,\rho$, and $\psi$ are the scalar Lagrange multipliers associated with constraints C1 and C3--C6 in (\ref{eqn:cross-layer-offline}), respectively.  $\boldmath\textbf{$\eta$}$  is the Lagrange multiplier vector  for subcarrier usage
constraint C8 and has elements $\eta_{i},i\in\{1,\ldots,n_F\}$.
Thus, the dual problem is given by
\begin{eqnarray}
\underset{\gamma,\psi,\rho,  \boldmath\textbf{$\eta$},\mu\ge
0}{\min}\ \underset{{\cal P}_j, {\cal
 S}_j}{\max}\,\,
{\cal L}(\gamma,\psi,\rho,  \boldmath\textbf{$\eta$},\mu,\nu,{\cal P}_j, {\cal
 S}_j).\label{eqn:master_problem-online}
\end{eqnarray}

\subsubsection*{Dual Decomposition and  Solution of Optimization Problem}

By using dual decomposition and following a similar approach as in (\ref{eqn:dual_problem})-(\ref{eqn:Final_power_circuit_2}),
 the resource allocation policy can be obtained via an iterative approach. For a  given set of Lagrange multipliers $\{\gamma,\,\psi,\,\rho$,\,$\boldmath\textbf{$\eta$}$,\,$\mu \}$,
 the power allocation ${\cal P}_j^*=\{\tilde{P}_{i,k}^{E*}[j],\tilde{P}_{i,k}^{N*}[j],{P}_{C}^{E*}[j],{P}_{C}^{N*}[j]\}$ and the subcarrier allocation ${\cal S}_j^*=\{s_{i,k}^*[j]\}$ for dual problem (\ref{eqn:master_problem-online}) for the signals from the BS to user $k$ in subcarrier $i$ in epoch $j$
are given by
\begin{eqnarray}\label{eqn:Final_power_allocation1-suboptimal}
\tilde{P}_{i,k}^{E*}[j]&=& s_{i,k}[j]{P}_{i,k}^{E*}[j]= s_{i,k}[j]\Bigg[\frac{W(\alpha_k+\rho)}{(\ln(2)(q \phi\varepsilon+\psi+\gamma))}
-\frac{1}{\Gamma_{i,k}[j]}\Bigg]^+\,\,\,\\
 \tilde{P}_{i,k}^{N*}[j]&=& s_{i,k}[j]{P}_{i,k}^{N*}[j]= s_{i,k}[j]\Bigg[\frac{W(\alpha_k+\rho)}{(\ln(2)(q\varepsilon+\mu\varepsilon+\psi))}
-\frac{1}{\Gamma_{i,k}[j]}-\tilde{P}_{i,k}^{E*}[j]\Bigg]^+, \label{eqn:Final_power_allocation2-suboptimal}  \end{eqnarray}\begin{eqnarray}
\label{eqn:sub_selection-suboptimal}s_{i,k}^*[j]&=&
 \left\{ \begin{array}{rl}
 1 &\mbox{if $k=\arg \underset{c}{\max} \,\, \ Q_{i,c}[j]$ }\\
 0 &\mbox{ otherwise}
       \end{array} \right., \\
\label{eqn:Final_power_circuit_1-suboptimal}
{P}_{C}^{E*}[j] &=&\Bigg[\frac{E[j]-\overline{L_E}\varepsilon \tilde{P}_{i,k}^{E*}[j]}{\overline{L_E}}\Bigg]_0^{P_C}, \,\mbox{and}\,\,
 {P}_{C}^{N*}[j]=P_C-{P}_{C}^{E*}[j],\label{eqn:Final_power_circuit_2-suboptimal}
 \end{eqnarray}
 where $\phi< 1$ and $Q_{i,k}[j]$ is defined in (\ref{eqn:subcarrier_allocation}). It can be observed that the proposed suboptimal online solution (\ref{eqn:Final_power_allocation1-suboptimal})--(\ref{eqn:Final_power_circuit_2-suboptimal}) shares some common properties with the asymptotically optimal offline solution in (\ref{eqn:Final_power_allocation1})--(\ref{eqn:Final_power_circuit_2}). In particular, the BS will prefer to first consume energy from the energy harvester  for $\phi<1$. If the energy provided by the energy harvester is not sufficient for achieving the maximum weighted energy efficiency of the system, the BS will start to consume energy from the non-renewable energy source. However, here, the value of $q$ in (\ref{eqn:Final_power_allocation1-suboptimal}) and (\ref{eqn:Final_power_allocation2-suboptimal}) is calculated w.r.t. the average epoch length $\overline{L_E}$. On the contrary,   the value of $q$ in (\ref{eqn:cap_discrete}) captures the effects of all channel gains and  energy arrivals in the time horizon of $T$ seconds.

We can update the set of Lagrange multipliers
$\{\gamma,\,\psi,\,\rho$,\,$\mu\}$   for a given ${{\cal P}_j, {\cal S}_j}$ by using the gradient method, since
the dual function is differentiable. The gradient update equations
are given by:
\begin{eqnarray}\label{eqn:multipler1-online}
\hspace*{-5mm}\gamma(\varsigma+1)\hspace*{-3mm}&=&\hspace*{-3mm}\Big[\gamma(\varsigma)-\xi_1(\varsigma)\hspace*{-1mm} \times  \hspace*{-1mm}
\Big(E[j]-\sum_{i=1}^{n_F}\sum_{k=1}^K \overline{L_E}\varepsilon \tilde{P}_{i,k}^E[j]-\overline{L_E} P_C^E[j]\Big)\Big]^+,\\
\label{eqn:multipler2-online}
\hspace*{-5mm} \mu(\varsigma+1)\hspace*{-3mm}&=&\hspace*{-3mm}\Big[\mu(\varsigma)-\xi_2(\varsigma)\hspace*{-1mm} \times  \hspace*{-1mm}
\Big(\overline{L_E} P_N -\sum_{i=1}^{n_F}\sum_{k=1}^K \overline{L_E}\varepsilon \tilde{P}_{i,k}^N[j]-\overline{L_E} P_C^N[j]\Big)\hspace*{-1mm}\Big]^+,\\
\hspace*{-5mm}\rho(\varsigma+1)\hspace*{-3mm}&=&\hspace*{-3mm}\Big[\rho(\varsigma)-\xi_3(\varsigma)\hspace*{-1mm} \times  \hspace*{-1mm}
\Big(\sum_{k=1}^K
\sum_{i=1}^{n_F}\sum_{j=1}^{L}  \overline{L_E} C_{i,k}[j] -\frac{R_{\min}}{Z}\Big)\Big]^+,
\label{eqn:multipler3-online}\\
\hspace*{-5mm}\psi(\varsigma+1)\hspace*{-3mm}&=&\hspace*{-3mm}\Big[\psi(\varsigma)-\xi_4(\varsigma)\hspace*{-1mm} \times  \hspace*{-1mm}
\Big(\overline{L_E} P_{\max} -\Big(\sum_{i=1}^{n_F}\sum_{k=1}^K \overline{L_E} \tilde{P}_{i,k}[j]  \Big)\Big)\Big]^+.
\label{eqn:multipler4-online}
\end{eqnarray}
Similar to the case of offline algorithm design, updating $\boldmath\textbf{$\eta$}$ is not necessary since it will not affect the subcarrier allocation in (\ref{eqn:sub_selection-suboptimal}).
A summary of the overall algorithm is given in Table I. In
each iteration of the main loop, we solve (\ref{eqn:lag_offline}) in line 7 of
Algorithm 1 for a given parameter $q$ via dual decomposition,
cf. (\ref{eqn:Final_power_allocation1-suboptimal})-(\ref{eqn:multipler4-online}). Then, we update parameter $q$ and use it for
solving (\ref{eqn:lag_offline}) in the next iteration. This
procedure is repeated until the proposed algorithm converges.

We now analyze the complexity of the proposed suboptimal online algorithm.
The proposed iterative algorithm requires the execution of
two nested loops in each event. The complexity of the outer loop, i.e., Algorithm 1, can be proved to be
linear in $n_F$ \cite{JR:convergence}. On the other hand, the inner loop
optimization problem in (\ref{eqn:lag_offline}) is  jointly concave w.r.t. the optimization variables. As a result, the solution for the problem formulation in (\ref{eqn:master_problem-online}) can be obtained with
a  complexity quadratically  in each epoch for the worst case, i.e., the complexity is
${\cal O}(n_F \times K^2)$ where $K^2$ is due to the worst case complexity in calculating (\ref{eqn:sub_selection-suboptimal}). As a result, the complexity of the proposed algorithm for an average of $Z$ events in $T$ seconds is ${\cal O}(  n_F \times  K^2\times Z)$.

\vspace*{-2mm}

\section{Results and Discussions}
\label{sect:5} In this section, we evaluate the
system performance for the proposed resource allocation and
scheduling algorithms using simulations. A micro-cell system with radius 500 m is considered. The number of subcarriers is $n_F=128$ with carrier
center frequency $2.5$ GHz, system bandwidth ${\cal B}=5$ MHz,
and $\alpha_k=1,\forall k$.   Each subcarrier  has
a bandwidth of $39$ kHz and the noise variance is
$\sigma_z^2=-128$ dBm. The 3GPP urban path loss model is used \cite{3Gpp:09}  with a reference distance of
$d_0=35$ m.   The $K$ desired users are uniformly distributed
between the reference distance and the cell boundary. The small
scale fading coefficients of the BS-to-user links are generated as
independent and identically distributed (i.i.d.) Rayleigh random
variables.  The multipath channel characteristic of each user is
 assumed to follow the power delay profile according of the  LTE extended pedestrian A channel model  \cite{book:Tap}. The  static circuit power
consumption is set to $P_C=$ 40 dBm \cite{CN:power_consumption_elements}. Unless specified
otherwise, the minimum data rate
requirement of the system is $R_{\min}=5$ Mbits/s.   We assume a transmission duration of $T=10$ seconds. The maximum transmit power  allowance $P_{\max}$ will be specified in each case study.  The energy harvester has a maximum energy storage of $E_{\max}=500$ J and an initial energy  $E_0=0$ J in the battery\footnote{The
values of the battery capacity and energy arrival (/harvesting) rates used in the paper are for
illustration purpose. In practice, the choice of battery capacity
should scale with the energy arrival (/harvesting) rates and the cell size. }. The amount of energy that can be harvested by the energy harvester in each energy epoch  is assumed to be a fixed amount of $5$ J. Then, the energy harvesting rate of the system is $5\lambda_E$ Joule/s.  The value of $\phi$ is set to  $\phi=0.01$ to account for the preference for harvested energy. The conference time for the multipath fading coefficients of each fading block is $200$  ms. We set $\epsilon=0.01$ for calculating the \emph{optimal online resource allocation policy} via DP. Furthermore, we
assume a power efficiency of $35\%$ in the PA, i.e.,
$\varepsilon=\frac{1}{0.35}= 2.8571$.  The average weighted system
energy efficiency is obtained by counting the number of weighted bits
which are successfully decoded by the receiver over the total energy
consumption averaged over the small scale fading. Note that if the resource allocator
is unable to guarantee the minimum data rate $R_{\min}$ in $T$, we
set the weighted energy efficiency and the average system capacity for these channel realizations to zero to account for the corresponding failure. Besides, unless further specified,  in the
following results, the ``number of iterations'' refers to the
number of iterations of Algorithm 1 in Table I.
\begin{figure}[t]
\centering\vspace*{-1cm}
\includegraphics[width=4.5in]{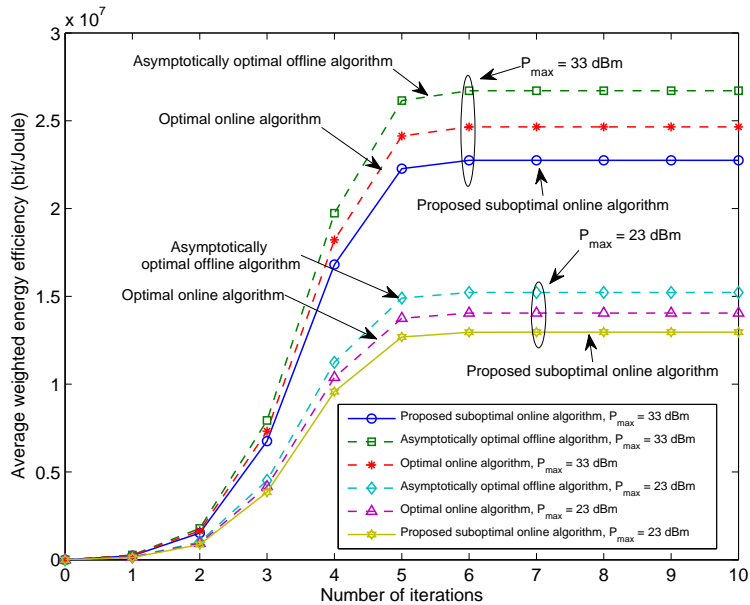}\vspace*{-0.5cm}
\caption{Average weighted energy efficiency (bit-per-Joule) versus
number of iterations with  different maximum transmit power
allowances, $P_{\max}$, for $K=5$ users,  a maximum power supplied by the non-renewable source $P_N=50$ dBm, and an energy harvesting rate of  $20$ Joule/s.  }\vspace*{-1cm}\label{figure:q_convergence}
\end{figure}
\vspace*{-2mm}

\subsection{Convergence of Proposed Iterative Algorithm }
Figure \ref{figure:q_convergence} illustrates the evolution of the
proposed suboptimal online iterative algorithm for different  maximum transmit power allowances, $P_{\max}$, $K=5$ users, and an energy harvesting rate of $20$ Joule/s. The results in Figure
\ref{figure:q_convergence} were averaged over $10^5$ independent
adaptation processes where each adaptation process involves
a different realization of the path loss and the small scale fading.
It can be observed that on average, in each case, the suboptimal iterative algorithm converges to above 83\% and 90\% of the weighted energy efficiency of the asymptotically \emph{optimal offline}  and \emph{optimal online} algorithms within 5 iterations, respectively. On the other hand, the inner loop for solving (\ref{eqn:cross-layer-offline}) converges within 5 iterations in each event. In other words, on average
the overall algorithm requires in total around $5\times 5\times Z$ iterations (inner loops and outer loops in $T$ seconds) to converge where $Z$ is the average number of events during $T$ seconds.

In the following case studies, we  set the number of iterations in the proposed suboptimal algorithm to 5.

\subsection{Energy Efficiency and Average Capacity versus Energy Harvesting Rates}
\begin{figure}[t]\centering\vspace*{-1cm}
\includegraphics[width=4.5in]{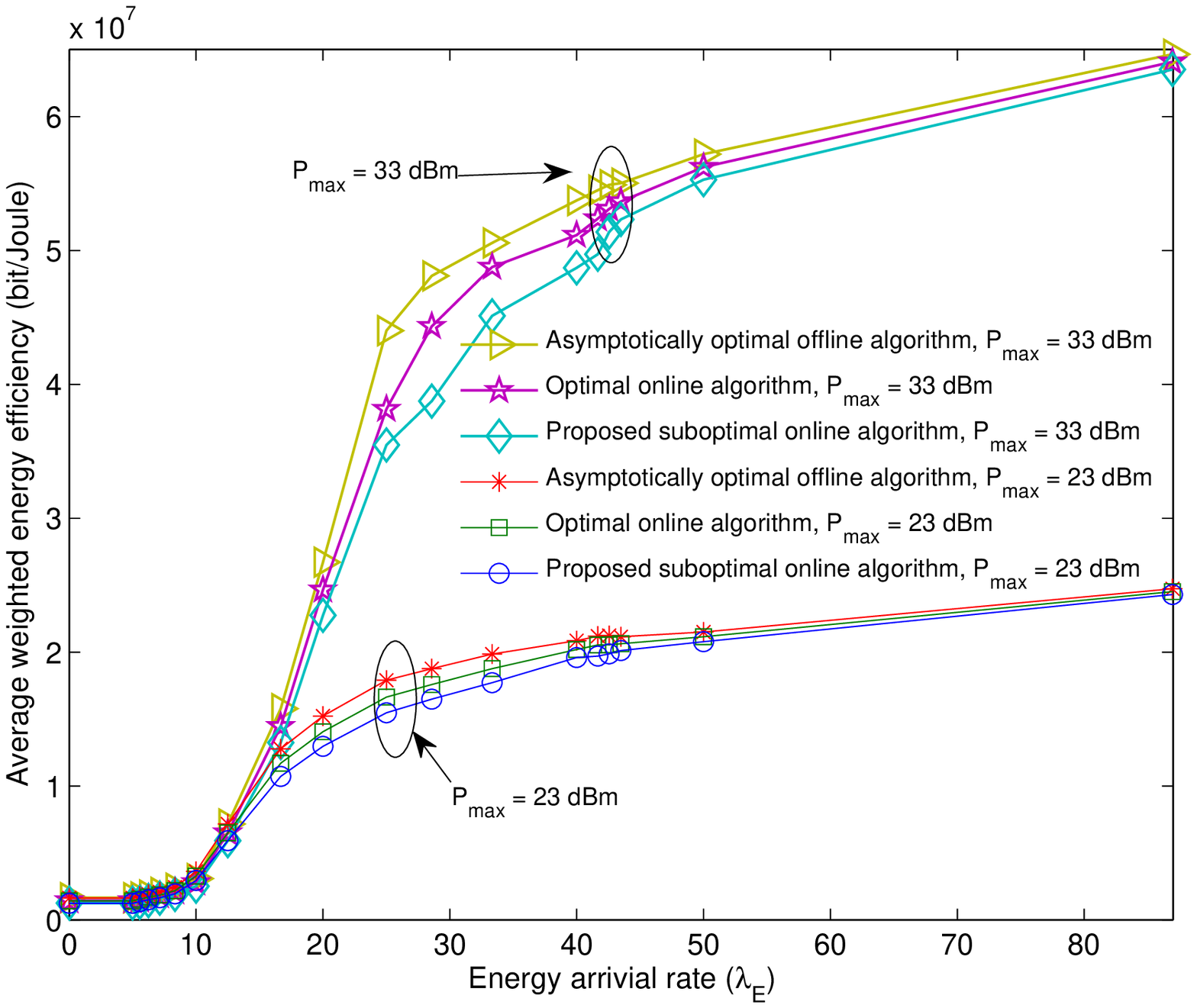}\vspace*{-0.5cm}
\caption{Average weighted energy efficiency (bit-per-Joule) versus energy harvesting rate (Joule-per-second)
for the proposed suboptimal algorithm and the benchmark schemes for different values of maximum transmit power allowance, $P_{\max}$.  The maximum power supplied by the non-renewable source is set to $P_N=50$ dBm. } \label{figure:ee_arrivial}
\includegraphics[width=4.5in]{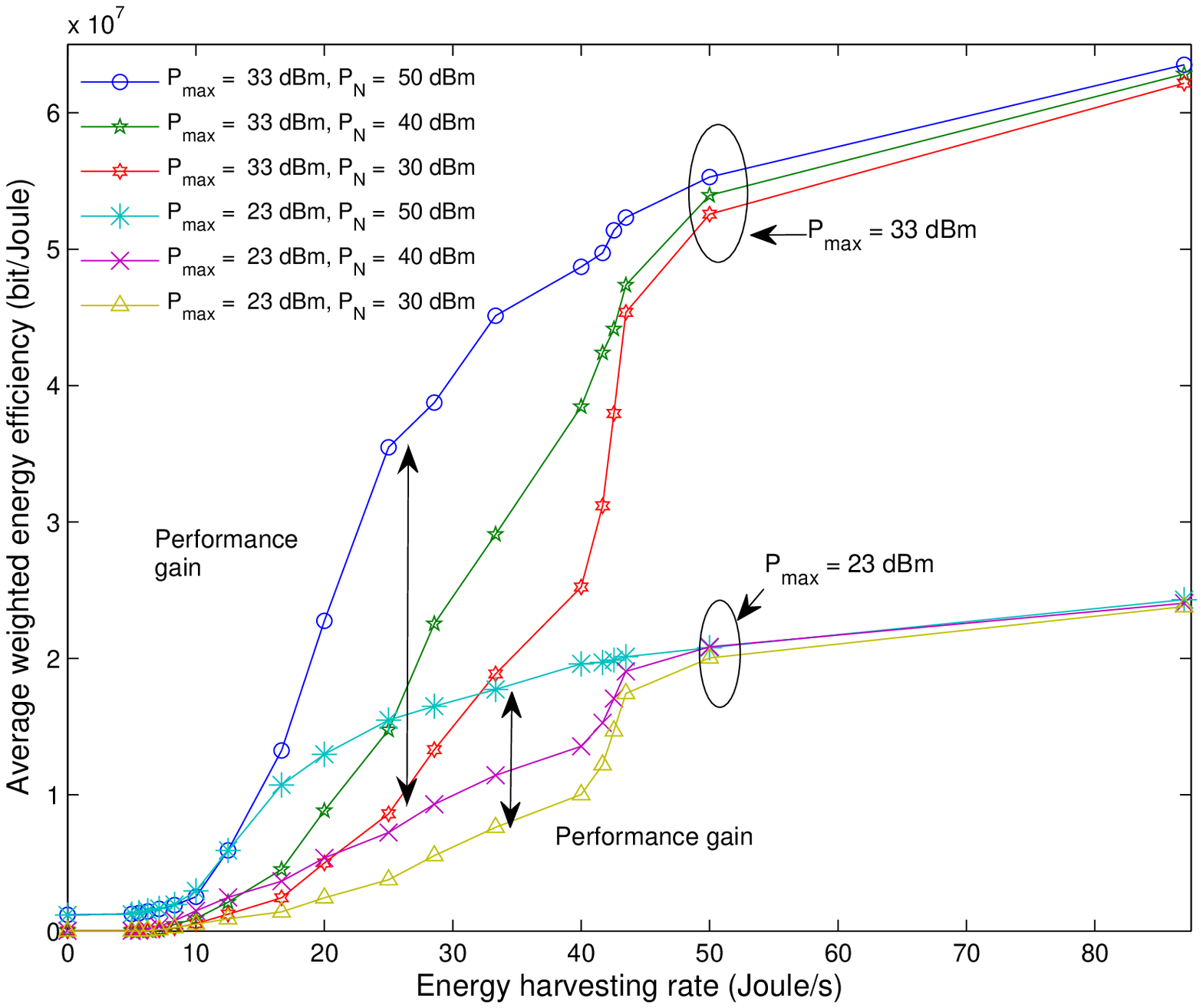}\vspace*{-0.5cm}
\caption{Average weighted energy efficiency (bit-per-Joule) versus energy harvesting rate (Joule-per-second)
for the proposed suboptimal algorithm for different values for maximum  non-renewable energy supply, $P_N$, and maximum transmit power allowance, $P_{\max}$. The double sided arrows represent the performance gains due to larger values of $P_N$. }\label{figure:ee_arrivial_different_PN}\vspace*{-0.4cm}\end{figure}

Figure \ref{figure:ee_arrivial} illustrates the average weighted energy efficiency versus the energy harvesting rate, for different
maximum transmit power allowances, $P_{\max}$, and $K=5$
users. 
 It can be observed that
the average weighted energy efficiency  of the proposed suboptimal algorithm  increases rapidly for increasing energy harvesting rate. This is because more  energy is available at  the energy harvester  as the  energy harvesting rate increases. As a result, the resource allocator  reduces its reliance  on the energy supplied by the non-renewable energy source by exploiting a larger amount of energy from the energy harvester.  For comparison, Figure
\ref{figure:ee_arrivial} also contains results for both the \emph{optimal online} algorithm and the asymptotically \emph{optimal offline} algorithm which serve as  performance benchmarks. It can be observed
that the proposed suboptimal algorithm  has a  performance close to that of the benchmark algorithms in all considered scenarios. In particular, the performance of the proposed suboptimal algorithm  approaches the benchmark schemes in both the low  and the high energy harvesting rate regimes. This is because in the low energy harvesting rate regime, the energy supplied by  the energy harvester is very limited. Thus, the BS has to rely mainly on the non-renewable source for maintaining normal operation and the influence of the energy harvester on  system performance becomes insignificant.  In the other extreme, the high energy harvesting rate converts the energy harvester into  a continuous energy source. As a result,  knowledge  about the future  energy arrivals in the asymptotically optimal offline algorithm becomes less valuable for resource allocation purpose, since there is always sufficient harvested energy for system operation in each epoch.

\begin{figure}[t]\centering\vspace*{-1cm}
\includegraphics[width=4.5in]{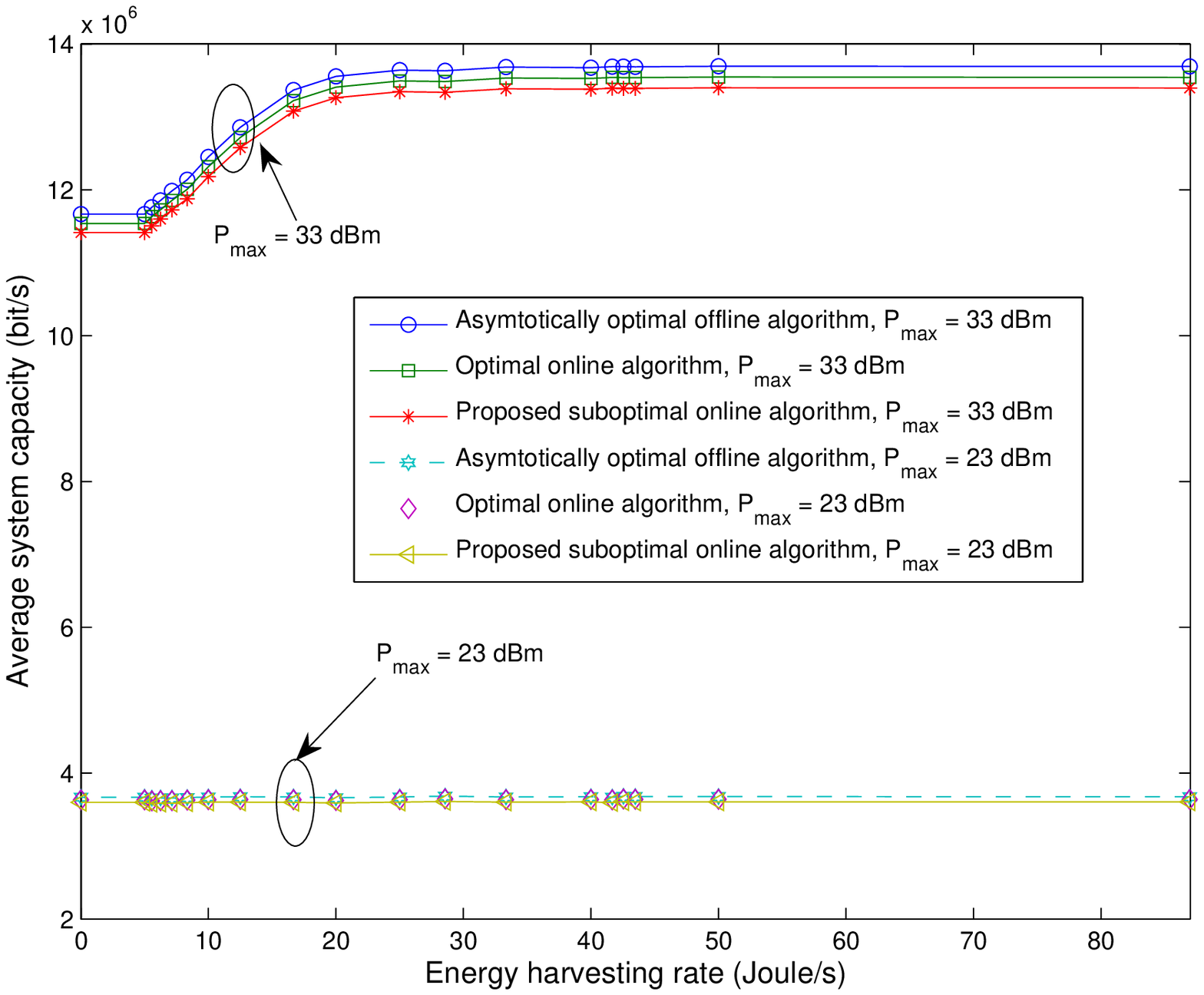}\vspace*{-0.2cm}
\caption{Average system capacity (bit/s) versus energy harvesting rate (Joule-per-second) for
the proposed suboptimal algorithm and the benchmark schemes for different values of maximum transmit power allowance, $P_{\max}$.  The maximum power supplied by the non-renewable source is set to $P_N=50$ dBm.  }\label{figure:cap_arrivial}\vspace*{-0.5cm}
\end{figure}

 Figure \ref{figure:ee_arrivial_different_PN} depicts the average weighted energy efficiency (bit-per-Joule) versus energy harvesting rate
for the proposed suboptimal algorithm for different values of maximum  non-renewable energy supply $P_N$ and maximum transmit power allowances $P_{\max}$. Figure  \ref{figure:ee_arrivial_different_PN} provides useful insight for system design as far as  the choice of the  maximum output power for the non-renewable energy supply is concerned. It can be observed that for all considered scenarios, a higher value of $P_N$ achieves a better average weighted energy efficiency. This is because a larger value of $P_N$ allows a higher flexibility in resource allocation since the non-renewable energy can be used as a supplement for the harvested energy whenever there is insufficient energy in the battery. However, there is a diminishing return in performance as $P_N$ increases in the high energy harvesting rate regime. This is due to the fact that for a large value of energy harvesting rate, the BS is able to consume a large amount of energy from the energy harvester which reduces the dependence on the non-renewable energy source. As a result, a small output power of the non-renewable energy supply is only preferable when the energy harvester is able to harvest a large amount of energy.

\begin{figure}[t]\centering
\includegraphics[width=4.5in]{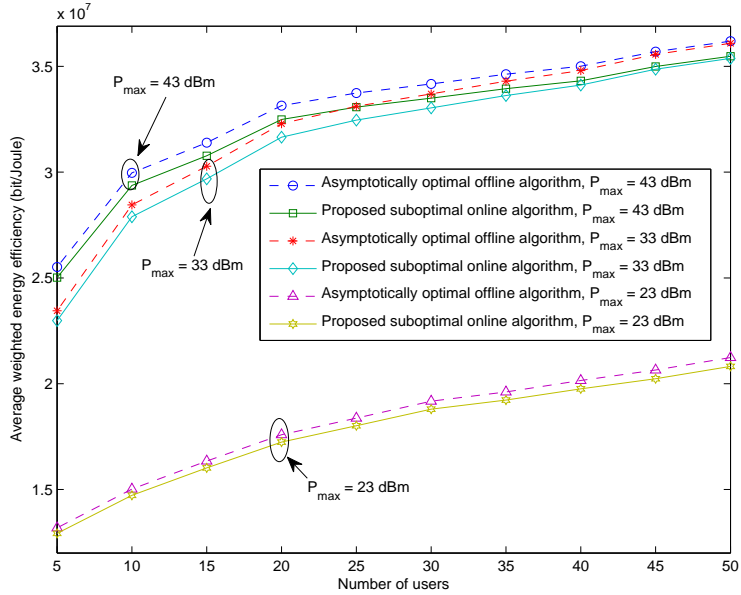}\vspace*{-0.5cm}
\caption{Average weighted energy efficiency (bit-per-Joule) versus the number of users $K$ for the proposed suboptimal algorithm and the asymptotically optimal offline scheme for different values of maximum transmit power allowance, $P_{\max}$, and an energy harvesting rate of $20$ Joule/s.  The maximum power supplied by the non-renewable source is set to $P_N=50$ dBm.  } \label{figure:ee_k}\vspace*{-0.5cm}
\end{figure}

Figure \ref{figure:cap_arrivial} shows the average system capacity versus energy harvesting rate for $K=5$
users
 and different maximum transmit power allowances $P_{\max}$. We compare the system performance
 of the proposed suboptimal algorithm again with the two aforementioned benchmark schemes.
 It can be observed that the average system capacity of the
proposed suboptimal algorithm approaches a constant in the high energy harvesting rate regime for the case of  $P_{\max}= 33$  dBm. This is because the proposed suboptimal algorithm clips the transmit power at the BS in order to maximize
the weighted system energy efficiency. However, when the maximum transmit power allowance is
small, i.e., $P_{\max} = 23$ dBm,   the system capacity performance gains due to  a high energy harvesting rate are quickly saturated for all schemes since the system capacity is always limited by the
small amount of radiated power in the RF.  On the other hand, we note that,
as expected, the benchmark schemes achieve a higher average system
capacity than the proposed suboptimal online algorithm,  since the proposed suboptimal
scheme utilizes only  the CSI of the current epoch.


\vspace*{-2mm}
\subsection{Energy Efficiency versus Number of Users}

Figures \ref{figure:ee_k}  depicts the weighted  energy
efficiency  versus the number of
users. Different
maximum transmit power allowances $P_{\max}$ at the BS are assumed for an energy harvesting rate of $20$ Joule/s. Note that the performance of the optimal online algorithm is not shown here  since the computational complexity in solving (\ref{eqn:dp}) becomes prohibitive for a large number of users $K$.
  It can be
observed that in all considered cases, the performances of the proposed suboptimal online algorithm and the asymptotically optimal offline algorithm scale with the number of users with a similar slope. In other words, the proposed suboptimal online algorithm is able to exploit multiuser diversity (MUD) for enhancing the system performance. Indeed,  MUD introduces
an extra power/energy gain \cite[Chapter
 6.6]{book:david_wirelss_com} to the system which facilities further
 energy savings. On the other hand, there is a diminishing return in the weighted energy efficiency for increasing the maximum transmit power allowance from
$P_{\max}=33$ dBm to $P_{\max}=43$ dBm, since both schemes  are not willing to consume exceedingly large amounts of energy for signal transmission.

\vspace*{-2mm}
\section{Conclusions}\label{sect:conclusion}
In this paper, we formulated the resource allocation algorithm
design for OFDMA systems with  hybrid energy harvesting BSs  as a non-convex optimization problem, in which the
circuit energy consumption, the finite battery storage capacity, and a minimum system data rate
requirement were taken into consideration. We first studied the structure of the asymptotically optimal offline
resource allocation algorithm by assuming non-causal channel gain and energy arrival knowledge.   Then, the derived offline solution served as a building block for the design of a practical close-to-optimal online resource allocation algorithm requiring only causal system knowledge.
 Simulation results did not only unveil   the achievable
 maximum weighted energy efficiency, but showed also  that the proposed suboptimal online algorithm achieves a close-to-optimal
performance within a small number of iterations.   Interesting topics for future work include studying the effects of imperfect CSI and  energy leakage.

\section*{appendix-Proof of Lemma \ref{lem:optimal_hold}}
\label{sect:appendix}
 \setcounter{subsection}{0}

The proof of Lemma  \ref{lem:optimal_hold} is divided into two parts. In the first part, we prove the concavity of the optimization problem in (\ref{eqn:inner_loop}). Then, in the second part, we prove a necessary condition for the optimal resource allocation policy based on the result in part one.

\vspace*{-2mm}
\subsection{Proof of the Concavity of the Transformed Problem in (\ref{eqn:inner_loop})} \label{appendix:concave}
We first consider the concavity of the objective function on a per
subcarrier basis w.r.t. all optimization variables. For the sake of notational simplicity, we define the receive channel gain-to-noise ratio (CNR) and the channel capacity between the BS and user $k$ on subcarrier $i$ at time instant $t$ as $\Gamma_{i,k}(t)=\frac{|H_{i,k}(t)|^2g_k(t)}{N_0W}$ and  $C_{i,k}(t)=s_{i,k}(t)W\log_2(1+\frac{\tilde {P}_{i,k}(t)\Gamma_{i,k}(t)}{s_{i,k}(t)})$, respectively. Let the objective function in
(\ref{eqn:inner_loop}) on subcarrier $i$ for user $k$ at time instant $t$ be
$f_{i,k}(t, {\cal P}, {\cal S})=\alpha_kC_{i,k}(t)-q\big[\varepsilon \phi\tilde {P}_{i,k}^E(t)+\varepsilon \tilde {P}_{i,k}^N(t)+ \phi {P}_{C}^E(t)+ {P}_{C}^N(t)\big]$. Then, we define $\mathbf{H}(f_{i,k}(t, {\cal P}, {\cal S}))$ and
$\varphi_1$, $\varphi_2$, \ldots, and  $\varphi_5$ as the Hessian matrix of function $f_{i,k}(t, {\cal P}, {\cal S})$
and the five eigenvalues of $\mathbf{H}(f_{i,k}(t, {\cal P}, {\cal S}))$, respectively. The
Hessian  of function $f_{i,k}(t, {\cal P}, {\cal S})$ and the corresponding eigenvalues are given by
%
\begin{eqnarray}\hspace*{-8mm}\mathbf{H}(f_{i,k}(t, {\cal P}, {\cal S}))&=&
\renewcommand\arraystretch{0.6}
\begin{bmatrix}
  \Lambda_{i,k}(t) &   \Lambda_{i,k}(t) & \Psi_{i,k}(t)&0&0 \\
    \Lambda_{i,k}(t) & \Lambda_{i,k}(t) & \Psi_{i,k}(t)& 0 & 0 \\
     \Psi_{i,k}(t) &     \Psi_{i,k}(t)& \Upsilon_{i,k}(t)& 0 & 0\\
     0&0&0&0&0\\
0&0&0&0&0
\end{bmatrix},\varphi_1=\varphi_2=\varphi_3=\varphi_4=0,\, \notag\\
\hspace*{-8mm} \mbox{and } \varphi_5&=&\frac{-\Gamma_{i,k}^2(t)(W\alpha_k)[(\tilde{P}_{i,k}^E(t)+\tilde{P}_{i,k}^N(t))^2+2 s_{i,k}^2(t)]/\ln(2)/s_{i,k}(t)}{(s_{i,k}(t)+\Gamma_{i,k}(t)[\tilde{P}_{i,k}^E(t)+\tilde{P}_{i,k}^N(t)])^2}\le 0,
\end{eqnarray}
respectively, where $\Lambda_{i,k}(t)=\frac{-\Gamma_{i,k}^2(t)(W\alpha_k)s_{i,k}(t)/\ln(2)}{(s_{i,k}(t)+\Gamma_{i,k}(t)[\tilde{P}_{i,k}^E(t)+\tilde{P}_{i,k}^N(t)])^2}$ ,
$\Psi_{i,k}(t)= \frac{-\Gamma_{i,k}^2(t)(W\alpha_k)s_{i,k}(t)/\ln(2)}{(s_{i,k}(t)+\Gamma_{i,k}(t)[\tilde{P}_{i,k}^E(t)+\tilde{P}_{i,k}^N(t)])^2}$, and $\Upsilon_{i,k}(t)=\frac{-\Gamma_{i,k}^2(t)(W\alpha_k)(\tilde{P}_{i,k}^E(t)+\tilde{P}_{i,k}^N(t))/\ln(2)/s_{i,k}(t)}{(s_{i,k}(t)+\Gamma_{i,k}(t)[\tilde{P}_{i,k}^E(t)+\tilde{P}_{i,k}^N(t)])^2}$. Hence, $\mathbf{H}(f_{i,k}(t, {\cal P}, {\cal S}))$ is a negative semi-definite
matrix since $\varphi_{\varrho}\le 0, \varrho=1,\ldots, 5 $. Therefore, $f_{i,k}(t, {\cal P}, {\cal S})$ is jointly concave w.r.t.
 optimization variables $\tilde{P}_{i,k}^E(t)$, $\tilde{P}_{i,k}^N(t)$, ${P}_{C}^E(t)$, ${P}_{C}^N(t)$, and $s_{i,k}(t)$ at time instant $t$.
 Besides, the integration of $f_{i,k}(t, {\cal P}, {\cal S})$ over $t$
and  the sum of $f_{i,k}(t, {\cal P}, {\cal S})$ over indices $k$ and
$i$ preserve the concavity of the objective function in
(\ref{eqn:inner_loop}) \cite{book:convex}. On the other hand,
constraints C1-C9 in (\ref{eqn:inner_loop}) span a convex feasible set,
and thus the transformed problem is a concave optimization problem.

\subsection{Optimality of Constant Resource Allocation Policy in each Epoch}
 Without loss of generality, we consider the time interval $[t_1, t_2)$ of \emph{epoch} $1$ and time instant $\tau_1$, where $t_1\le \tau_1< t_2$. Suppose an adaptive resource allocation policy is adopted in $t_1\le \tau_1< t_2$ such that two constant resource allocation policies, $\{{\cal P}_1,  {\cal S}_1\}$ and $\{{\cal P}_2,  {\cal S}_2\}$, are applied in $t_1 \le t < \tau_1$ and $\tau_1 \le t < t_2$, respectively. We assume that  $\{{\cal P}_1,  {\cal S}_1\}$ and $\{{\cal P}_2,  {\cal S}_2\}$ are feasible solutions to (\ref{eqn:inner_loop}) while ${\cal P}_1\neq{\cal P}_2$ and ${\cal S}_1\neq {\cal S}_2$. Now, we define a third resource allocation policy $\{{\cal P}_3, {\cal S}_3 \}$ such that ${\cal P}_3=\frac{{\cal P}_1(\tau_1-t_1)+{\cal P}_2(t_2-\tau_1)}{t_2-t_1}$ and  ${\cal S}_3=\frac{{\cal S}_1(\tau_1-t_1)+{\cal S}_2(t_2-\tau_1)}{t_2-t_1}$. Note that arithmetic operations between any two resource allocation policies are defined element-wise.  Then, we apply resource allocation policy\footnote{Resource allocation policy  $\{{\cal P}_3, {\cal S}_3\}$ is also a feasible solution to (\ref{eqn:inner_loop}) by the convexity of the feasible solution set. } $\{{\cal P}_3, {\cal S}_3 \}$ to the entire \emph{epoch 1} and integrate $f_{i,k}(t, {\cal P}, {\cal S})$ over time interval $[t_1, t_2)$ which yields:
 \begin{eqnarray} \label{eqn:constant_proof}\notag
&&\hspace*{2mm}\int_{t_1}^{t_2} \sum_{i=1}^{n_F}\sum_{k=1}^K f_{i,k}(t, {\cal P}_3, {\cal S}_3)\, dt\\
 &\stackrel{(a)}{\ge}&\notag  \int_{t_1}^{t_2}\Big( \frac{\tau_1-t_1}{t_2-t_1}\sum_{i=1}^{n_F}\sum_{k=1}^Kf_{i,k}(t, {\cal P}_1, {\cal S}_1)+ \frac{t_2-\tau_1}{t_2-t_1}\sum_{i=1}^{n_F}\sum_{k=1}^Kf_{i,k}(t, {\cal P}_2, {\cal S}_2)\,\Big) dt  \\
&=&(\tau_1-t_1)\sum_{i=1}^{n_F}\sum_{k=1}^K f_{i,k}(t, {\cal P}_1, {\cal S}_1)+ (t_2-\tau_1) \sum_{i=1}^{n_F}\sum_{k=1}^K f_{i,k}(t, {\cal P}_2, {\cal S}_2) \notag\\&=&\int_{t_1}^{\tau_1} \sum_{i=1}^{n_F}\sum_{k=1}^K f_{i,k}(t, {\cal P}_1, {\cal S}_1)\, dt +\int_{\tau_1}^{t_2} \sum_{i=1}^{n_F}\sum_{k=1}^K f_{i,k}(t, {\cal P}_2, {\cal S}_2)\, dt,
 \end{eqnarray}
 where $(a)$  is due to the concavity of $f_{i,k}(t, {\cal P}, {\cal S})$ which was proved in the first part. In other words, for any non-constant resource allocation policy within an epoch, there always exists at least one constant resource allocation policy which achieves at least the same performance. As a result, the optimal resource allocation policy is constant within each epoch. \hspace*{8.0cm}$\square$
\begin{Remark}
We would like to emphasize that although the fact that the  resource allocation policy is constant within each epoch seems obvious in hindsight, it does not necessarily always hold. In the extreme case, when the transformed objective function is strictly convex w.r.t. to the optimization variables,  then we can replace ``${\ge}$"  by ``${<}$" in (\ref{eqn:constant_proof}) of the above proof. As a result, there always exists at least one adaptive resource allocation policy which outperforms the constant resource allocation policy.
\end{Remark}

\vspace*{-0.2cm}

\bibliographystyle{IEEEtran}
\bibliography{OFDMA-AF}

\makeatletter
\let\@oddfoot\@empty
\makeatother

\end{document}